\documentclass[
]{jss}

\usepackage{orcidlink,thumbpdf,lmodern}

\usepackage[utf8]{inputenc}

\author{
Nikos I. Bosse\\London School of Hygiene \& Tropical Medicine (LSHTM)
\AND Hugo Gruson\\LSHTM \And Anne Cori\\Imperial College London
\AND Edwin van Leeuwen\\UK Health Security Agency,
LSHTM \And Sebastian Funk\\LSHTM \And Sam Abbott\\LSHTM
}
\title{Evaluating Forecasts with \pkg{scoringutils} in \proglang{R}}

\Plainauthor{Nikos I. Bosse, Hugo Gruson, Anne Cori, Edwin van
Leeuwen, Sebastian Funk, Sam Abbott}
\Plaintitle{Evaluating Forecasts with scoringutils in R}
\Shorttitle{Evaluating Forecasts with \pkg{scoringutils} in
\proglang{R}}

\Abstract{
Evaluating forecasts is essential to understand and improve forecasting
and make forecasts useful to decision makers. A variety of \proglang{R}
packages provide a broad variety of scoring rules, visualisations and
diagnostic tools. One particular challenge, which \pkg{scoringutils}
aims to address, is handling the complexity of evaluating and comparing
forecasts from several forecasters across multiple dimensions such as
time, space, and different types of targets. \pkg{scoringutils} extends
the existing landscape by offering a convenient and flexible
\pkg{data.table}-based framework for evaluating and comparing
probabilistic forecasts (forecasts represented by a full predictive
distribution). Notably, \pkg{scoringutils} is the first package to offer
extensive support for probabilistic forecasts in the form of predictive
quantiles, a format that is currently used by several infectious disease
Forecast Hubs. The package is easily extendable, meaning that users can
supply their own scoring rules or extend existing classes to handle new
types of forecasts. \pkg{scoringutils} provides broad functionality to
check the data and diagnose issues, to visualise forecasts and missing
data, to transform data before scoring, to handle missing forecasts, to
aggregate scores, and to visualise the results of the evaluation. The
paper presents the package and its core functionality and illustrates
common workflows using example data of forecasts for COVID-19 cases and
deaths submitted to the European COVID-19 Forecast Hub.
}

\Plainkeywords{forecasting, forecast evaluation, proper scoring
rules, scoring, R}

\Address{
    Nikos I. Bosse\\
    London School of Hygiene \& Tropical Medicine (LSHTM)\\
    Centre for Mathematical Modelling of Infectious Diseases\\
London School of Hygiene \& Tropical Medicine\\
Keppel Street\\
London WC1E 7HT\\
  E-mail: \email{nikos.bosse@lshtm.ac.uk}\\
  URL: \url{https://lshtm.ac.uk}\\~\\
      Hugo Gruson\\
    LSHTM\\
    Centre for Mathematical Modelling of Infectious Diseases\\
London School of Hygiene \& Tropical Medicine\\
Keppel Street\\
London WC1E 7HT\\
  E-mail: \email{hugo.gruson@lshtm.ac.uk}\\
  
      Anne Cori\\
    Imperial College London\\
    MRC Centre for Global Infectious Disease Analysis, School of Public
Health\\
Imperial College London\\
Norfolk Place\\
London W2 1PG\\
  E-mail: \email{a.cori@imperial.ac.uk}\\
  
      Edwin van Leeuwen\\
    UK Health Security Agency, LSHTM\\
    Statistics, Modelling and Economics Department\\
UK Health Security Agency\\
London NW9 5EQ\\
  E-mail: \email{Edwin.VanLeeuwen@phe.gov.uk}\\
  
      Sebastian Funk\\
    LSHTM\\
    Centre for Mathematical Modelling of Infectious Diseases\\
London School of Hygiene \& Tropical Medicine\\
Keppel Street\\
London WC1E 7HT\\
  E-mail: \email{sebastian.funk@lshtm.ac.uk}\\
  
      Sam Abbott\\
    LSHTM\\
    Centre for Mathematical Modelling of Infectious Diseases\\
London School of Hygiene \& Tropical Medicine\\
Keppel Street\\
London WC1E 7HT\\
  E-mail: \email{sam.abbott@lshtm.ac.uk}\\
  
  }

\providecommand{\tightlist}{%
  \setlength{\itemsep}{0pt}\setlength{\parskip}{0pt}}

\usepackage{booktabs}
\usepackage{longtable}
\usepackage{array}
\usepackage{multirow}
\usepackage{wrapfig}
\usepackage{float}
\usepackage{colortbl}
\usepackage{pdflscape}
\usepackage{tabu}
\usepackage{threeparttable}
\usepackage{threeparttablex}
\usepackage[normalem]{ulem}
\usepackage{makecell}
\usepackage{xcolor}

\usepackage{amsmath}

\shortcites{reichCollaborativeMultiyearMultimodel2019, kukkonenReviewOperationalRegionalscale2012, funkShorttermForecastsInform2020, cramerEvaluationIndividualEnsemble2021, bracherShorttermForecastingCOVID192021, sherrattPredictivePerformanceMultimodel2022, bracherNationalSubnationalShortterm2022, cramerCOVID19ForecastHub2020}

\usepackage{amssymb} \usepackage{caption}

 \newcommand{\fct}[1]{\code{#1()}} \usepackage{booktabs} \usepackage{multirow} \usepackage{makecell} \usepackage{graphicx}

\begin{document}

\section{Introduction}\label{introduction}

Good forecasts are of great interest to decision makers in various
fields like finance
\citep{timmermannForecastingMethodsFinance2018, elliottForecastingEconomicsFinance2016},
weather predictions
\citep{gneitingWeatherForecastingEnsemble2005, kukkonenReviewOperationalRegionalscale2012}
or infectious disease modeling
\citep{reichCollaborativeMultiyearMultimodel2019, funkShorttermForecastsInform2020, cramerEvaluationIndividualEnsemble2021, bracherNationalSubnationalShortterm2022, sherrattPredictivePerformanceMultimodel2022}.
For decades, researchers, especially in the field of weather
forecasting, have therefore developed and refined an arsenal of
techniques to evaluate predictions (see for example
\cite{goodRationalDecisions1952},
\cite{epsteinScoringSystemProbability1969, murphyNoteRankedProbability1971a, mathesonScoringRulesContinuous1976},
\cite{gneitingProbabilisticForecastsCalibration2007},
\cite{funkAssessingPerformanceRealtime2019},
\cite{gneitingStrictlyProperScoring2007},
\cite{bracherEvaluatingEpidemicForecasts2021}).

Various \proglang{R} \citep{R} packages cover a wide variety of scoring
rules, plots and metrics that are useful in assessing the quality of a
forecast. Existing packages offer functionality that is well suited to
evaluate a variety of predictive tasks, but also come with important
limitations.

Some packages such as \pkg{tscount} \citep{tscount}, \pkg{topmodels}
\citep{topmodels}, \pkg{GLMMadaptive} \citep{GLMMadaptive}, \pkg{cvGEE}
\citep{cvGEE} or \pkg{fabletools} \citep{fabletools} expect that
forecasts were generated in a certain way and require users to supply an
object of a specific class to compute scores. These packages provide
excellent tools for users operating within the specific package
framework but are by their nature not generally applicable to many use
cases practitioners might encounter.

Packages such as \pkg{scoringRules} \citep{scoringRules}, \pkg{Metrics}
\citep{Metrics}, \pkg{MLmetrics} \citep{MLmetrics}, \pkg{verification}
\citep{verification}, \pkg{SpecsVerification} \citep{SpecsVerification},
\pkg{surveillance} \citep{surveillance}, \pkg{predtools}
\citep{predtools}, or \pkg{probably} \citep{probably} provide an
extensive collection of tools, scoring rules and visualisations for
various use cases. However, most scoring functions operate on vectors
and matrices. This is desirable in many applications but can make it
difficult to simultaneously evaluate multiple forecasts across several
dimensions, such as time, space, and different types of targets.

\pkg{scoring} \citep{scoring} operates on a data.frame and uses a
formula interface, making this task easier. However, \pkg{scoring} only
exports a few scoring rules and does not allow users to supply their
own. \pkg{yardstick} \citep{yardstick}, which builds on the
\pkg{tidymodels} \citep{tidymodels} framework, is the most general and
flexible other forecast evaluation package. It allows users to apply
arbitrary scoring rules to a data.frame of forecasts, independently of
how they were created. However, \pkg{yardstick} is primarily focused on
point forecasts and classification tasks. It currently lacks general
support for probabilistic forecasts (forecasts in the form of a full
predictive distribution, represented e.g., by a set of quantiles or
samples from the forecast distribution). Probabilistic forecasts are
desirable, as they allow decision makers to take into account the
uncertainty of a forecast
\citep{gneitingProbabilisticForecastsCalibration2007}, and are widely
used, e.g., in Meteorology or Epidemiology.

\pkg{scoringutils} aims to fill the existing gap in the ecosystem by
providing a flexible general-purpose tool for the evaluation of
probabilistic forecasts. It offers a coherent \code{data.table}-based
framework and workflow that allows users to evaluate and compare
forecasts across multiple dimensions using a wide variety of default and
user-provided scoring rules. Notably, \pkg{scoringutils} is the first
package to offer extensive support for probabilistic forecasts in the
form of predictive quantiles, a format that is currently used by several
infectious disease Forecast Hubs
\citep{reichCollaborativeMultiyearMultimodel2019, cramerCOVID19ForecastHub2020, sherrattPredictivePerformanceMultimodel2022, bracherNationalSubnationalShortterm2022}.
The package provides broad functionality to check the data and diagnose
issues, to visualise forecasts and missing data, to transform data
before scoring \citep[see][]{bosseScoringEpidemiologicalForecasts2023},
to apply various metrics and scoring rules to data, to handle missing
forecasts, to aggregate scores and to visualise the results of the
evaluation. \pkg{scoringutils} makes extensive use of \pkg{data.table}
\citep{data.table} to ensure fast and memory-efficient computations. The
core functionality is designed around S3 classes, allowing users to
expand on the generics and methods implemented in the package.
\pkg{scoringutils} provides extensive documentation and case studies, as
well as sensible defaults for scoring forecasts.

\begin{CodeChunk}
\begin{figure}[!h]

{\centering \includegraphics[width=1\linewidth]{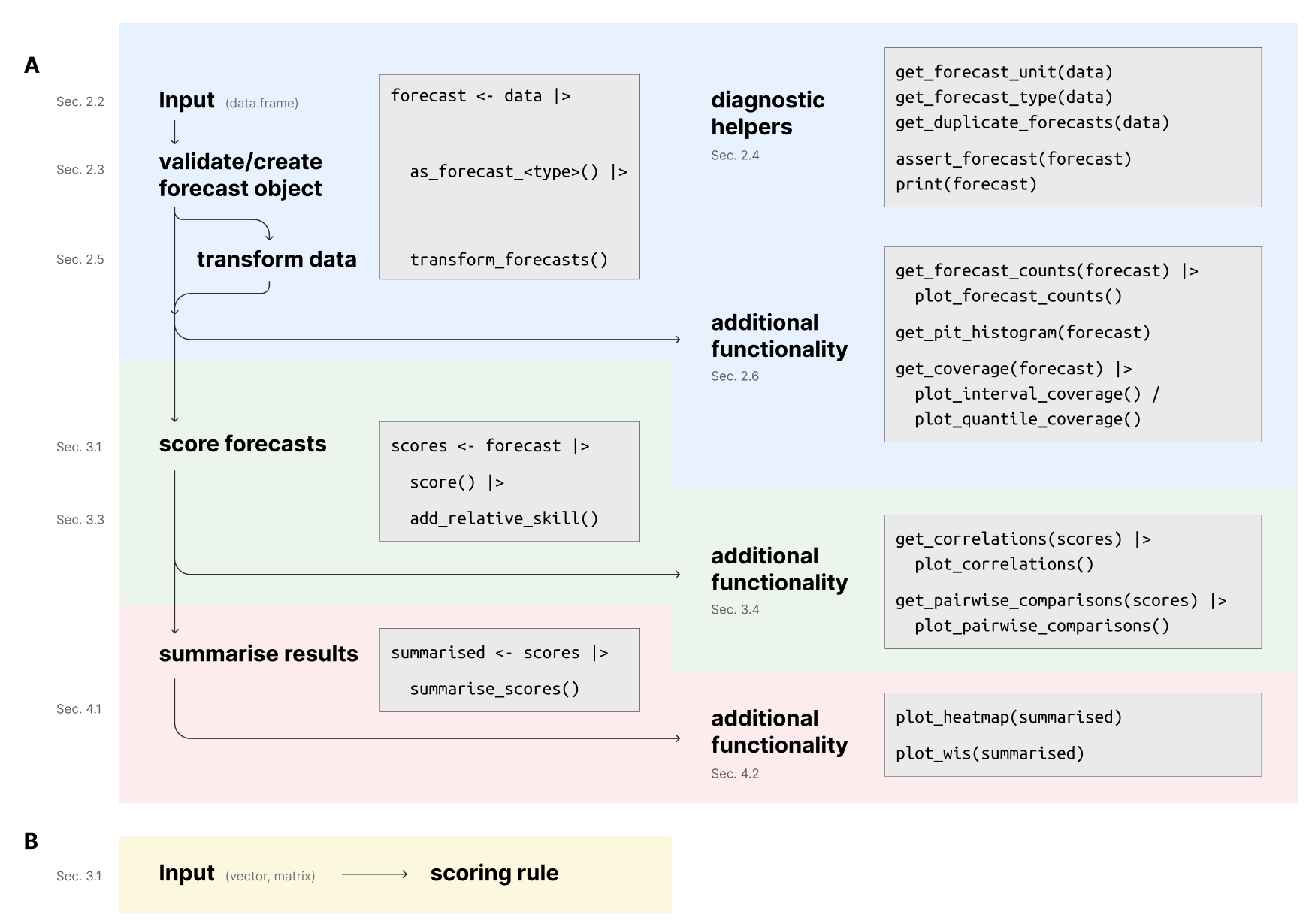} 

}

\caption{Illustration of the suggested workflow for evaluating forecasts with \pkg{scoringutils}. A: Workflow for working with forecasts in a \code{data.table}-based format. The left side shows the core workflow of the package: 1) validating and processing inputs, 2) scoring forecasts and 3) summarising scores. The right side shows additional functionality that is available at the different stages of the evaluation process. The part in blue is covered by Section \ref{sec:inputs} and includes all functions related to processing and validating inputs as well as obtaining additional information about the forecasts. The part in green is covered by Section \ref{sec:scoring} and includes all functions related to scoring forecasts and obtaining additional information about the scores. The part in red is covered by Section \ref{sec:summarising} and includes all functions related to summarising scores and additional visualisations based on summarised scores. B: An alternative workflow, allowing users to call scoring rules directly with vectors/matrices as inputs.}\label{fig:workflow-scoringutils}
\end{figure}
\end{CodeChunk}

\subsubsection{Paper outline and package
workflow}\label{paper-outline-and-package-workflow}

The structure of this paper follows the suggested package workflow which
consists of 1) validating and processing inputs, 2) scoring forecasts
and 3) summarising scores. This workflow is illustrated in Figure
\ref{fig:workflow-scoringutils}, which displays the core workflow (left
side) as well as additional functionality that is available at different
stages of the evaluation process (right side).

Section \ref{sec:inputs} is centred around validating inputs,
\code{forecast} objects, and the associated functionality. It explains
the expected input formats and how to validate inputs and diagnose
issues. It provides an overview of the types of forecasts supported by
\pkg{scoringutils} and the different S3 classes used to represent these
forecast types. It also provides information on a variety of functions
that can be used to visualise forecasts, transform inputs or obtain
additional information and visualisations.

Section \ref{sec:scoring} is centred around scoring forecasts and the
additional functionality that is available to manipulate and analyse
scores further. It explains how to score forecasts, either in a
\code{data.table}-format or in a format based on matrices and vectors.
It also provides information on additional information that can be
computed from scores, such as correlations between scores or relative
skill scores based on pairwise comparisons. These can be useful to
mitigate the effects of missing forecasts.

Section \ref{sec:summarising} is centred around summarised scores. It
explains how to summarise scores and gives information on additional
visualisations that can be created based on summarised scores.

Section \ref{sec:discussion} discusses the merits and limitations of the
package in its current version as explores avenues for future work.

All functionality will be illustrated using the example data shipped
with the package, which is based on a subset of case and death forecasts
submitted every week between May and September 2021 to the European
COVID-19 Forecast Hub
\citep{sherrattPredictivePerformanceMultimodel2022}. Following the
convention of the different COVID-19 Forecast Hubs, we will restrict
examples to two-week-ahead forecasts.

The code for this package and paper can be found on
\url{https:github.com/epiforecasts/scoringutils}. The full package
documentation as well as an overview of all existing functions can also
be seen on \url{https://epiforecasts.io/scoringutils}.

\section{Inputs, forecast types and input validation} \label{sec:inputs}

\subsection{Input formats and types of
forecasts}\label{input-formats-and-types-of-forecasts}

Forecasts differ in the exact prediction task and in how the forecaster
chooses to represent their prediction. To distinguish different kinds of
forecasts, we use the term ``forecast type'' (which is more a convenient
classification than a formal definition). Currently,
\texttt{scoringutils} distinguishes five different forecast types:
``point'', ``binary'', ``nominal'', ``quantile'' and ``sample''
forecasts.

\begin{itemize}
\tightlist
\item
  ``Point'' denotes a forecast for a continuous or discrete outcome
  variable that is represented by a single number.
\item
  ``Binary'' denotes a probability forecast for a binary (yes/no)
  outcome variable. This is sometimes also called ``soft binary
  classification''.
\item
  ``Nominal'' denotes a probability forecast for a variable where the
  outcome can assume one of multiple unordered classes. This represents
  a generalisation of binary forecasts to multiple possible outcomes.
\item
  ``Quantile'' or ``quantile-based'' is used to denote a probabilistic
  forecast for a continuous or discrete outcome variable, with the
  forecast distribution represented by a set of predictive quantiles.
  While a single quantile would already satisfy the requirements for a
  quantile-based forecast, most scoring rules expect a set of quantiles
  which are symmetric around the median (thus forming the lower and
  upper bounds of central ``prediction intervals'') and will error (or
  return \texttt{NA} if \texttt{na.rm\ =\ TRUE}) if this is not the
  case.
\item
  ``Sample'' or ``sample-based'' is used to denote a probabilistic
  forecast for a continuous or discrete outcome variable, with the
  forecast represented by a finite set of samples drawn from the
  predictive distribution. A single sample technically suffices, but
  would lead to very imprecise results.
\end{itemize}

\begin{table}[h]
\centering
\resizebox{\textwidth}{!}{
\setlength\tabcolsep{10pt}
\begin{tabular}{@{}lllll@{}} % Define the table with five columns
\toprule
\textbf{Forecast type} & & & \textbf{column} & \textbf{type} \\
\midrule
% Classification
\multirow{5}{*}{\makecell[cl]{Categorical\\forecast}}    & \multirow{2}{*}{Binary}     & Soft classification & \texttt{observed} & factor with 2 levels \\ 
                                &                             & {\footnotesize(prediction is probability)}  & \texttt{predicted} & numeric [0,1] \\
\cmidrule(l){2-5} 
                                & \multirow{3}{*}{\makecell[cl]{Nominal\\{\footnotesize(multiclass)}}} & \multirow{3}{*}{\makecell[cl]{Soft classification\\{\footnotesize(prediction is probability)}}} 
                                                                  & \texttt{observed}   & factor with $N$ levels \\ 
                                &                                 & & \texttt{predicted} & numeric [0,1] \\
                                &                                 & & \texttt{predicted\_label} & factor with $N$ levels \\
\midrule

% Point forecasts
\multirow{2}{*}{Point forecast} & & & \texttt{observed}  & numeric                \\
                                 & & & \texttt{predicted} & numeric                \\
\midrule

% Probabilistic forecast
\multirow{6}{*}{\makecell[cl]{Probabilistic\\forecast}} & & \multirow{3}{*}{Sample format} & \texttt{observed} & numeric                \\
                                                         &                      &               & \texttt{predicted} & numeric                \\
                                                         &                      &               & \texttt{sample\_id} & numeric              \\
\cmidrule(l){3-5} 
                                                         &                      & \multirow{3}{*}{Quantile format}  & \texttt{observed} & numeric                \\
                                                         &                      &               & \texttt{predicted} & numeric                \\
                                                         &                      &               & \texttt{quantile\_level} & numeric [0,1]  \\
\bottomrule
\end{tabular}
}
\caption{Formatting requirements for data inputs. For binary forecasts, the column \texttt{observed} must be of type factor with two levels and the column \texttt{predicted} must be a numeric between 0 and 1. For nominal forecasts, the observed value must be a factor with $N$ levels (where $N$ is the number of possible outcomes) and a column \texttt{predicted\_label} must denote the outcome for which a probability was made. For all other forecast types, both \texttt{observed} and \texttt{predicted} must be of type numeric. Forecasts in a sample-based format require an additional numeric column \texttt{sample\_id} and forecasts in a quantile-based format require an additional numeric column \texttt{quantile\_level} with values between 0 and 1.}
\label{tab:input-score}
\end{table}

The starting point for working with \pkg{scoringutils} is usually a
\code{data.frame} (or similar) containing both the predictions and the
observed values. In a next step (see Section \ref{sec:validation}) this
data will be validated and transformed into a ``forecast object'' (a
\code{data.table} with a class \texttt{forecast} and an additional class
corresponding to the forecast type). The input data needs to have a
column \texttt{observed} for the observed values, a column
\texttt{predicted} for the predicted values. Additional requirements
depend on the forecast type.

Table \ref{tab:input-score} shows the expected input format for each
forecast type.

The package contains example data for each forecast type, which can
serve as an orientation for the correct formats. The example data sets
are exported as \texttt{example\_point} and \texttt{example\_binary},
\texttt{example\_nominal}, \texttt{example\_quantile},
\texttt{example\_sample\_continuous}, and
\texttt{example\_sample\_discrete}. For illustrative purposes, the
example data also contains some rows with only observations and no
corresponding predictions. All example data in the package use a column
called \texttt{model} to denote the name of the model/forecaster that
generated the forecast. This is also the default in some functions, but
does not reflect a hard requirement. Input formats for the scoring rules
that can be called directly follow the same convention, with inputs
expected to be vectors or matrices.

\subsubsection{The unit of a single
forecast}\label{the-unit-of-a-single-forecast}

Apart from the columns \texttt{observed}, \texttt{predicted},
\texttt{model}, and the extra columns required for each forecast type,
it is usually necessary that the input data contains additional columns.
This is because a single probabilistic forecast (apart from binary
predictions) is composed of multiple values. A quantile-based forecast,
for example, is composed of several quantiles, and a sample-based
forecast of multiple samples. However, every row only holds a single
sample/quantile. Several rows in the input data therefore jointly form a
single forecast. Additional columns in the input provide the information
necessary to group rows that belong to the same forecast. The
combination of values in those columns forms the unit of a single
forecast (or ``forecast unit'') and should uniquely identify a single
forecast. For example, consider forecasts made by different models in
various locations at different time points and for different targets. A
single forecast could then be uniquely described by the values in the
columns \texttt{model}, \texttt{location}, \texttt{date}, and
\texttt{target}, and the forecast unit would be
\texttt{forecast\_unit\ =\ c("model",\ "location",\ "date",\ "target")}.

Rows are automatically grouped based on the values in all other columns
present in the data (excluding required columns like \texttt{sample\_id}
or \texttt{quantile\_level} and values computed by \pkg{scoringutils}).
As the forecast unit is determined based on all existing columns, no
column must be present that is unrelated to the forecast unit. As a very
simplistic example, consider an additional row, \texttt{"even"}, that is
one if the row number is even and zero otherwise. The existence of this
column would change results, as \pkg{scoringutils} assumes it was
relevant to grouping the forecasts.

\subsection{Forecast objects and input validation} \label{sec:validation}

The raw input data needs to be processed and validated by converting it
into a \texttt{forecast} object (ignore for now that the example data
shipped with package is pre-validated by default).

\begin{CodeChunk}
\begin{CodeInput}
R> library("scoringutils")
R> forecast_quantile <- example_quantile[horizon == 2] |>
+   as_forecast_quantile() 
\end{CodeInput}
\end{CodeChunk}

Every forecast type has a corresponding
\texttt{as\_forecast\_\textless{}type\textgreater{}()} function that
transforms the input into a \texttt{forecast} object and validates it. A
forecast object is a \texttt{data.table} that has passed some input
validations. It behaves like a \texttt{data.table}, but has an
additional class \texttt{forecast} as well as a class corresponding to
the forecast type (\texttt{forecast\_point}, \texttt{forecast\_binary},
\texttt{forecast\_nominal}, \texttt{forecast\_quantile} or
\texttt{forecast\_sample}).

All \texttt{as\_forecast\_\textless{}type\textgreater{}()} functions can
take additional arguments that help facilitate the process of creating a
forecast object:

\begin{CodeChunk}
\begin{CodeInput}
R> forecast_quantile <- example_quantile[horizon == 2] |>
+   as_forecast_quantile(
+     forecast_unit = c(
+       "model", "location", "target_end_date", 
+       "forecast_date", "horizon", "location"
+     ),
+     observed = "observed", 
+     predicted = "predicted",
+     model = "model",
+     quantile_level = "quantile_level",
+   ) 
\end{CodeInput}
\end{CodeChunk}

The argument \texttt{forecast\_unit} allows the user to manually set the
unit of a single forecast. This is done by dropping all columns that are
not either specified in the \texttt{forecast\_unit} or are ``protected''
columns (such as \texttt{observed}, \texttt{predicted}, \texttt{model},
\texttt{quantile\_level}, or \texttt{sample\_id}). The other arguments
can be used to specify the column names of the input data that
correspond to the required columns. The function will rename the
specified columns to the corresponding required columns.

\subsection{Diagnostic helper
functions}\label{diagnostic-helper-functions}

Various helper functions are available to diagnose and fix issues with
the input data. A simple one is the \texttt{print()} method for forecast
objects. Once a forecast object has successfully been created, the
forecast type and the forecast unit will automatically be added to the
output when printing.

\begin{CodeChunk}
\begin{CodeInput}
R> print(forecast_quantile, 2)
\end{CodeInput}
\begin{CodeOutput}
Forecast type: quantile
\end{CodeOutput}
\begin{CodeOutput}
Forecast unit:
\end{CodeOutput}
\begin{CodeOutput}
location, target_end_date, target_type, location_name, forecast_date,
model, and horizon
\end{CodeOutput}
\begin{CodeOutput}

Key: <location, target_end_date, target_type>
      location target_end_date target_type observed location_name
        <char>          <Date>      <char>    <num>        <char>
   1:       DE      2021-05-15       Cases    64985       Germany
   2:       DE      2021-05-15       Cases    64985       Germany
  ---                                                            
7014:       IT      2021-07-24      Deaths       78         Italy
7015:       IT      2021-07-24      Deaths       78         Italy
      forecast_date quantile_level predicted                 model
             <Date>          <num>     <int>                <char>
   1:    2021-05-03          0.010     63106 EuroCOVIDhub-ensemble
   2:    2021-05-03          0.025     67867 EuroCOVIDhub-ensemble
  ---                                                             
7014:    2021-07-12          0.975       611  epiforecasts-EpiNow2
7015:    2021-07-12          0.990       719  epiforecasts-EpiNow2
      horizon
        <num>
   1:       2
   2:       2
  ---        
7014:       2
7015:       2
\end{CodeOutput}
\end{CodeChunk}

Internally, the print method calls \code{get\_forecast\_type()} and
\code{get\_forecast\_unit()}. Both functions can also be called
independently. \code{get\_forecast\_type()} and
\code{get\_forecast\_unit()} work on either an unvalidated
\code{data.frame} (or similar) or on an already validated forecast
object. They return the forecast type and the forecast unit,
respectively, as inferred from the input data.

\code{assert\_forecast()} asserts that an existing forecast object
passes all validations and returns \texttt{invisble(NULL)} if the
forecast object is valid (and otherwise errors).

One common issue that causes transformation to a \texttt{forecast}
object to fail are ``duplicates'' in the data. \pkg{scoringutils}
strictly requires that there be only one forecast per forecast unit and
only one predicted value per quantile level or sample id within a single
forecast. Duplicates usually occur if the forecast unit is misspecified.
For example, if we removed the column \texttt{target\_type} from the
example data, we would now have two forecasts (one for cases and one for
deaths of COVID-19) that appear to have the same forecast unit (since
the information that distinguished between case and death forecasts is
no longer there). The function \code{get\_duplicate\_forecasts()}
returns duplicate rows for the user to inspect. To remedy the issue, the
user needs to add additional columns that uniquely identify a single
forecast.

\begin{CodeChunk}
\begin{CodeInput}
R> rbind(example_quantile, example_quantile[1001:1002]) |>
+   get_duplicate_forecasts() 
\end{CodeInput}
\begin{CodeOutput}
   location target_end_date target_type observed location_name
     <char>          <Date>      <char>    <num>        <char>
1:       DE      2021-05-22      Deaths     1285       Germany
2:       DE      2021-05-22      Deaths     1285       Germany
3:       DE      2021-05-22      Deaths     1285       Germany
4:       DE      2021-05-22      Deaths     1285       Germany
   forecast_date quantile_level predicted                model
          <Date>          <num>     <int>               <char>
1:    2021-05-17          0.975      1642 epiforecasts-EpiNow2
2:    2021-05-17          0.975      1642 epiforecasts-EpiNow2
3:    2021-05-17          0.990      1951 epiforecasts-EpiNow2
4:    2021-05-17          0.990      1951 epiforecasts-EpiNow2
   horizon
     <num>
1:       1
2:       1
3:       1
4:       1
\end{CodeOutput}
\end{CodeChunk}

\subsection{Transforming forecasts}\label{transforming-forecasts}

As suggested in \cite{bosseScoringEpidemiologicalForecasts2023}, users
may want to transform forecasts before scoring them. Two commonly used
scoring rules are the continuous ranked probability score (CRPS) and the
weighted interval score (WIS). Both measure the absolute distance
between the forecast and the observation. This may not be desirable, for
example in the context of epidemiological forecasts, where infectious
disease processes are usually modelled to occur on a multiplicative
scale. Taking the logarithm of the forecasts and observations before
scoring them makes it possible to evaluate forecasters based on how well
they predicted the exponential growth rate.

The function \code{transform\_forecasts()} takes a validated forecast
object as input and allows users to apply arbitrary transformations to
forecasts and observations. Users can specify a function via the
argument \code{fun} (as well as supply additional function parameters).
The default function is \code{log_shift()}, which is simply a wrapper
around \code{log()} with an additional argument that allows adding an
offset (i.e., \code{log(x + offset)}) to deal with zeroes in the data.
Users can specify to either append the transformed forecasts to the
existing data by setting \code{append = TRUE} (the default behaviour,
resulting in an additional column \texttt{scale}) or to replace the
existing forecasts in place.

The example data contains negative values which need to be handled
before applying the logarithm. Presumably, negative values for count
data should be dropped altogether, but for illustrative purposes, we
will call \code{transform\_forecasts()} twice to replace them with
zeroes first before appending transformed counts.

\begin{CodeChunk}
\begin{CodeInput}
R> forecast_quantile |> 
+   transform_forecasts(fun = \(x) {pmax(x, 0)}, append = FALSE) |>
+   transform_forecasts(fun = log_shift, offset = 1) |>
+   print(2)
\end{CodeInput}
\begin{CodeOutput}
Forecast type: quantile
\end{CodeOutput}
\begin{CodeOutput}
Forecast unit:
\end{CodeOutput}
\begin{CodeOutput}
location, target_end_date, target_type, location_name, forecast_date,
model, horizon, and scale
\end{CodeOutput}
\begin{CodeOutput}

       location target_end_date target_type     observed
         <char>          <Date>      <char>        <num>
    1:       DE      2021-05-15       Cases 64985.000000
    2:       DE      2021-05-15       Cases 64985.000000
   ---                                                  
14029:       IT      2021-07-24      Deaths     4.369448
14030:       IT      2021-07-24      Deaths     4.369448
       location_name forecast_date quantile_level    predicted
              <char>        <Date>          <num>        <num>
    1:       Germany    2021-05-03          0.010 63106.000000
    2:       Germany    2021-05-03          0.025 67867.000000
   ---                                                        
14029:         Italy    2021-07-12          0.975     6.416732
14030:         Italy    2021-07-12          0.990     6.579251
                       model horizon   scale
                      <char>   <num>  <char>
    1: EuroCOVIDhub-ensemble       2 natural
    2: EuroCOVIDhub-ensemble       2 natural
   ---                                      
14029:  epiforecasts-EpiNow2       2     log
14030:  epiforecasts-EpiNow2       2     log
\end{CodeOutput}
\end{CodeChunk}

\subsection{Additional functionality related to forecast
objects}\label{additional-functionality-related-to-forecast-objects}

\pkg{scoringutils} offers a variety of different functions that allow
users to obtain and visualise additional information about their
forecast. The package also has an extensive Vignette with examples for
further visualisations that are not implemented as functions.

\subsubsection{Displaying the number of forecasts
available}\label{displaying-the-number-of-forecasts-available}

Users can get an overview of how many forecasts there are using
\code{get\_forecast\_counts()}. The function takes a validated forecast
object as input and returns a data.table of forecast counts, which helps
obtain an overview of missing forecasts. This can impact the evaluation,
if missingness correlates with performance. Users can specify the level
of summary through the \texttt{by} argument. For example, to see how
many forecasts there are per \texttt{model}, \texttt{target\_type} and
\texttt{forecast\_date}, we can run

\begin{CodeChunk}
\begin{CodeInput}
R> forecast_counts <- forecast_quantile |>
+   get_forecast_counts(
+     by = c("model", "target_type", "forecast_date")
+   )
\end{CodeInput}
\end{CodeChunk}

We can visualise the results by calling \code{plot\_forecast\_counts()}
on the output (Figure \ref{fig:plot-forecast-counts}).

\begin{CodeChunk}
\begin{CodeInput}
R> library("ggplot2")
R> forecast_counts |>
+   plot_forecast_counts(x = "forecast_date") + 
+   facet_wrap(~ target_type) +
+   labs(y = "Model", x = "Forecast date")
\end{CodeInput}
\begin{figure}[!h]

{\centering \includegraphics[width=1\linewidth]{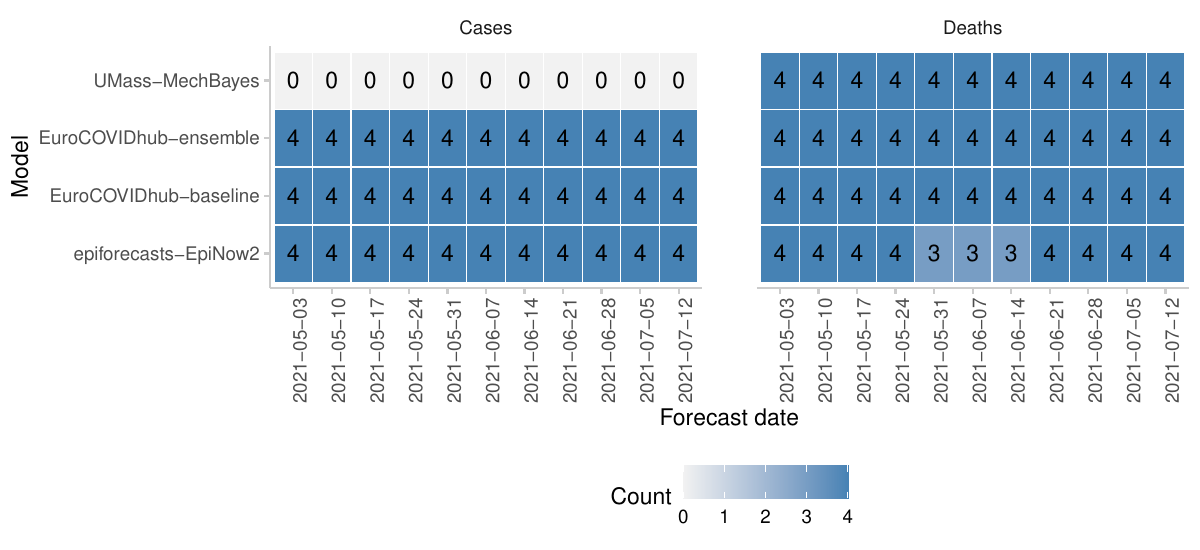} 

}

\caption[Visualistion of forecast counts for the example data]{Visualistion of forecast counts for the example data. Numbers (and colour shade) indicate the number of forecasts available for a given model, target type and forecast date.}\label{fig:plot-forecast-counts}
\end{figure}
\end{CodeChunk}

\subsubsection{Probabilistic calibration and PIT
histograms}\label{probabilistic-calibration-and-pit-histograms}

One important quality of good forecasts is calibration. The term
describes a statistical consistency between the forecasts and the
observations, i.e., an absence of systematic deviations between the two.
It is possible to distinguish several forms of calibration which are
discussed in detail by
\cite{gneitingProbabilisticForecastsCalibration2007}. The form of
calibration most commonly focused on is called probabilistic
calibration. Probabilistic calibration means that the forecast
distributions are consistent with the true data-generating distributions
in the sense that on average, \(\tau\)\% of true observations will be
below the corresponding \(\tau\)-\%-quantiles of the cumulative forecast
distributions.

A common way to visualise probabilistic calibration is the probability
integral transform (PIT) histogram
\citep{dawidPresentPositionPotential1984}. Observed values, \(y\), are
transformed using the CDF of the predictive distribution, \(F\), to
create a new variable \(u\) with \(u = F(y)\). \(u\) is therefore simply
the CDF of the predictive distribution evaluated at the observed value.
If forecasts are probabilistically calibrated, then the transformed
values will be uniformly distributed (for a proof see for example
\citet{angusProbabilityIntegralTransform1994}). When plotting a
histogram of PIT values (see Figure \ref{fig:pit-plots}), a systematic
bias usually leads to a triangular shape, a U-shaped histogram
corresponds to forecasts that are underdispersed (too sharp) and a hump
shape appears when forecasts are overdispersed (too wide). There exist
different variations of the PIT to deal with discrete instead of
continuous data (see e.g., \cite{czadoPredictiveModelAssessment2009} and
\cite{funkAssessingPerformanceRealtime2019}). The PIT version
implemented in \texttt{scoringutils} for discrete variables follows
\cite{funkAssessingPerformanceRealtime2019}.

Users can obtain PIT histograms based on validated forecast objects
using the function \code{get\_pit\_histogram()}. Once again, the
argument \texttt{by} controls the summary level. The output of the
following is shown in Figure \ref{fig:pit-plots}:

\begin{CodeChunk}
\begin{CodeInput}
R> example_sample_continuous |>
+   get_pit_histogram(by = c("model", "target_type")) |>
+   ggplot(aes(x = mid, y = density)) +
+   geom_col() +
+   facet_grid(target_type ~ model) +
+   labs(x = "Quantile", "Density")
\end{CodeInput}
\begin{figure}[!h]

{\centering \includegraphics[width=1\linewidth]{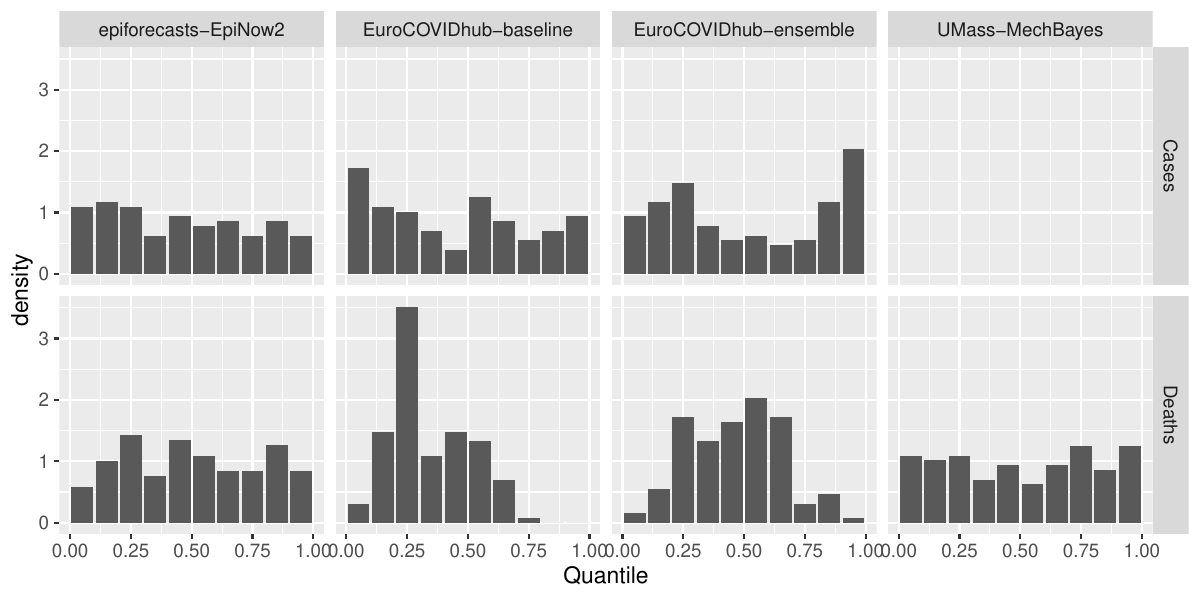} 

}

\caption[PIT histograms of all models stratified by forecast target]{PIT histograms of all models stratified by forecast target. Histograms should ideally be uniform. A u-shape usually indicates overconfidence (forecasts are too narrow), a hump-shaped form indicates underconfidence (forecasts are too uncertain) and a triangle-shape indicates bias.}\label{fig:pit-plots}
\end{figure}
\end{CodeChunk}

It is, in theory, possible to conduct a formal test for probabilistic
calibration, for example by employing an Anderson-Darling test on the
uniformity of PIT values. In practice, this can be difficult as
forecasts, and therefore PIT values as well, are often correlated.
Personal experience suggests that the Anderson-Darling test is often too
quick to reject the null hypothesis of uniformity. It is also important
to note that uniformity of the PIT histogram does not guarantee that
forecasts are indeed calibrated.
\cite{gneitingProbabilisticForecastsCalibration2007, hamillInterpretationRankHistograms2001a}
provide examples with different forecasters who are mis-calibrated, but
have uniform PIT histograms.

\subsubsection{Probabilistic calibration and coverage
plots}\label{probabilistic-calibration-and-coverage-plots}

For forecasts in a quantile-based format, there exists a second way to
assess probabilistic calibration: we can easily compare the proportion
of observations that fall below the \(\tau\)-quantiles of all forecasts
(``empirical quantile coverage'') to the nominal quantile coverage
\(\tau\). Similarly, we can compare the empirical coverage of the
central prediction intervals formed by the predictive quantiles to the
nominal interval coverage. For example, the central 50\% prediction
intervals of all forecasts should contain around 50\% of the observed
values, the 90\% central intervals should contain around 90\% of
observations etc. In addition, we can define coverage deviation as the
difference between nominal and empirical coverage.

Interval and quantile coverage can easily be computed by calling
\code{get_coverage()} on a validated forecast object (in a
quantile-based format). The function computes interval coverage,
quantile coverage, interval coverage deviation and quantile coverage
deviation and returns a \texttt{data.table} with corresponding columns.
Coverage values will be summarised according to the level specified in
the \texttt{by} argument and one value per quantile level/interval range
is returned.

\begin{CodeChunk}
\begin{CodeInput}
R> forecast_quantile |>
+   get_coverage(by = "model") |>
+   print(2)
\end{CodeInput}
\begin{CodeOutput}
                    model quantile_level interval_range
                   <char>          <num>          <num>
 1: EuroCOVIDhub-baseline           0.50              0
 2: EuroCOVIDhub-baseline           0.45             10
---                                                    
91:       UMass-MechBayes           0.01             98
92:       UMass-MechBayes           0.99             98
    interval_coverage interval_coverage_deviation quantile_coverage
                <num>                       <num>             <num>
 1:        0.00000000                 0.000000000         0.6818182
 2:        0.09090909                -0.009090909         0.6477273
---                                                                
91:        1.00000000                 0.020000000         0.0000000
92:        1.00000000                 0.020000000         1.0000000
    quantile_coverage_deviation
                          <num>
 1:                   0.1818182
 2:                   0.1977273
---                            
91:                  -0.0100000
92:                   0.0100000
\end{CodeOutput}
\end{CodeChunk}

Results can then be visualised using the functions
\texttt{plot\_interval\_coverage()} (see Figure \ref{fig:coverage}A) and
\texttt{plot\_quantile\_coverage()} (see \ref{fig:coverage}B). Both show
nominal against empirical coverage. Ideally, forecasters should lie on
the diagonal line. If the line moves into the green-shaded area, the
forecaster is too conservative, i.e., the predictive distributions are
too wide/overdispersed on average. The white area implies
overconfidence/predictive distributions that are too narrow on average
(see Figure \ref{fig:calibration-plots}) for more details).

\begin{CodeChunk}
\begin{CodeInput}
R> coverage <- get_coverage(forecast_quantile, by = c("model", "target_type")) 
R> 
R> plot_interval_coverage(coverage) + 
+   facet_wrap(~ target_type)
R> 
R> plot_quantile_coverage(coverage) + 
+   facet_wrap(~ target_type)
\end{CodeInput}
\end{CodeChunk}

Note that users can also compute individual coverage values as scores
using \code{score()}. This represents a separate workflow that allows
users to obtain coverage values as a summary measure to be computed
alongside other scores, rather than providing a way to visually assess
calibration.

\begin{CodeChunk}
\begin{figure}[!h]

{\centering \includegraphics[width=1\linewidth]{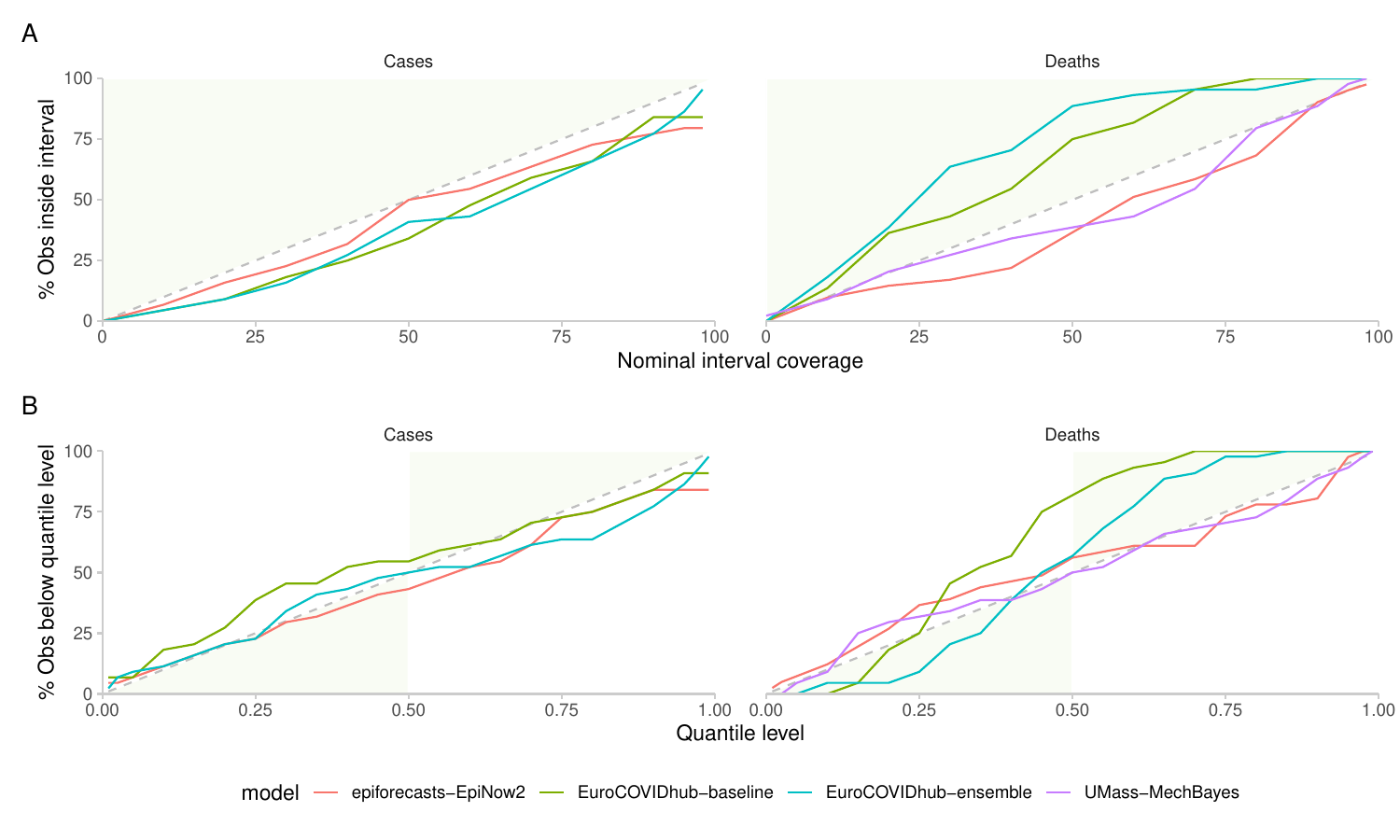} 

}

\caption[Interval coverage (A) and quantile coverage (B) plots]{Interval coverage (A) and quantile coverage (B) plots. Areas shaded in green indicate that the forecasts are too wide (i.e., underconfident), while areas in white indicate that the model is overconfident and generates too narrow prediction intervals.}\label{fig:coverage}
\end{figure}
\end{CodeChunk}

\section{Scoring forecasts} \label{sec:scoring}

Metrics and scoring rules can be applied to data in two different ways:
They can be conveniently applied to a data set of observed and predicted
values using \code{score()}, or they be called directly on a set of
vectors and matrices. This section will mostly focus on \code{score()}.

\subsection{score() and working with scoring
rules}\label{score-and-working-with-scoring-rules}

The function \code{score()} is the workhorse of the package and applies
a set of metrics and scoring rules to predicted and observed values. It
is a generic function that dispatches to different methods depending on
the class of the input. The input of \code{score()} is a validated
forecast object and its output is an object of class \texttt{scores},
which is a essentially \texttt{data.table} with an additional attribute
\texttt{metrics} (containing the names of the metrics used for scoring).

\begin{CodeChunk}
\begin{CodeInput}
R> example_point[horizon == 2] |>
+   score() |>
+   print(2)
\end{CodeInput}
\begin{CodeOutput}
Key: <location, target_end_date, target_type>
     location target_end_date target_type location_name forecast_date
       <char>          <Date>      <char>        <char>        <Date>
  1:       DE      2021-05-15       Cases       Germany    2021-05-03
  2:       DE      2021-05-15       Cases       Germany    2021-05-03
 ---                                                                 
304:       IT      2021-07-24      Deaths         Italy    2021-07-12
305:       IT      2021-07-24      Deaths         Italy    2021-07-12
                     model horizon ae_point   se_point       ape
                    <char>   <num>    <num>      <num>     <num>
  1: EuroCOVIDhub-ensemble       2    45731 2091324361 0.7037162
  2: EuroCOVIDhub-baseline       2    67622 4572734884 1.0405786
 ---                                                            
304:       UMass-MechBayes       2       46       2116 0.5897436
305:  epiforecasts-EpiNow2       2      108      11664 1.3846154
\end{CodeOutput}
\end{CodeChunk}

All \code{score()} methods take an argument \texttt{metrics} with a
named list of functions to apply to the data. These can be metrics
exported by \pkg{scoringutils} or any other custom scoring function. All
metrics scoring rules passed to \code{score()} need to adhere to the
same input format (see Figure \ref{fig:input-scoring-rules}),
corresponding to the type of forecast to be scored. Scoring functions
must accept a vector of observed values as their first argument, a
matrix/vector of predicted values as their second argument and, for
quantile-based forecasts, a vector of quantile levels as their third
argument). However, functions may have arbitrary argument names. Within
\code{score()}, inputs like the observed and predicted values, quantile
levels etc. are passed to the individual scoring rules by position,
rather than by name. The default scoring rules for point forecasts, for
example, comprise functions from the \pkg{Metrics} package, which use
the names \texttt{actual} and \texttt{predicted} for their arguments
instead of \texttt{observed} and \texttt{predicted}. Additional
arguments can be passed down to the scoring functions via the
\texttt{...} arguments in \code{score()}.

\begin{CodeChunk}
\begin{figure}[!h]

{\centering \includegraphics[width=1\linewidth]{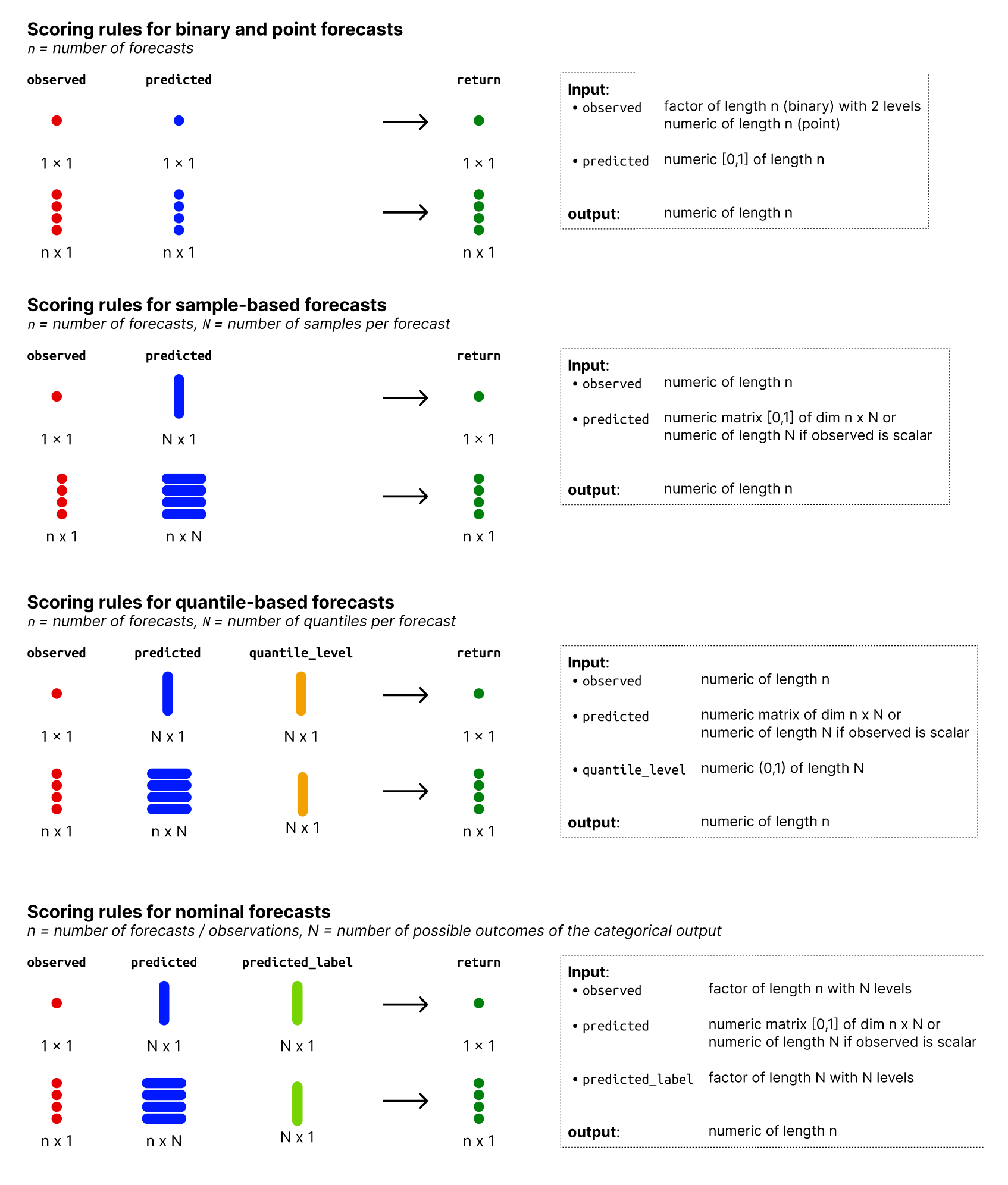} 

}

\caption{Overview of the inputs and outputs of the metrics and scoring rules exported by \pkg{scoringutils}. Dots indicate scalar values, while bars indicate vectors (comprised of values that belong together). Several bars (vectors) can be grouped into a matrix with rows representing the individual forecasts. All scoring functions used within \fct{score} must accept the same input formats as the functions here. However, functions used within \fct{score} do not necessarily have to have the same argument names (see Section \ref{sec:scoring}). Input formats directly correspond to the required columns for the different forecast types (see Table \ref{tab:input-score}). The only exception is the forecast type 'sample': Inputs require a column \code{sample\_id} in \fct{score}, but no corresponding argument is necessary when calling scoring rules directly on vectors or matrices.}\label{fig:input-scoring-rules}
\end{figure}
\end{CodeChunk}

\subsubsection{Composing a custom list of metrics and scoring
rules}\label{composing-a-custom-list-of-metrics-and-scoring-rules}

For every forecast type, there exists a default list of scoring rules
that are applied to the data when calling \code{score()}. The default
lists can be accessed by calling the function \code{get\_metrics()} on a
\texttt{forecast} object. \code{get\_metrics()} takes additional
arguments \texttt{exclude} and \texttt{select} which can be used to
customise which scoring rules are included. Alternatively, users can
call the function \code{select\_metrics()} on a list of scoring rules,
which achieves the same purposes and allows users to compose custom
lists of metrics and scoring rules.

\begin{CodeChunk}
\begin{CodeInput}
R> custom_metrics <- get_metrics(example_quantile) |>
+   select_metrics(select = c("wis", "overprediction"))
R> 
R> score(metrics = custom_metrics)
\end{CodeInput}
\end{CodeChunk}

\subsubsection{Details on metrics exported by scoringutils}

All metrics are named according to the following schema:
\texttt{\{metric\ name\}\_\{forecast\ type\}}. If only a single forecast
type is possible, then \texttt{\_\{forecast\ type\}} is omitted. The
return value is a vector with scores (only in the case of \code{wis()},
which is composed of three components (see \ref{sec:wis}), is there an
optional argument that causes the function to return a list of vectors
for the individual WIS components). The first argument of all metrics
exported by \pkg{scoringutils} is always \texttt{observed}, and the
second one is \texttt{predicted}. Scoring rules for quantile-based
forecasts have an additional argument, \texttt{quantile\_level}, to
denote the quantile levels of the predictive quantiles.

Metrics exported by \pkg{scoringutils} differ in the relationship
between input and output. Some scoring rules have a one-to-one
relationship between predicted values and scores, returning one value
per value in \texttt{predicted}. This is the case for all metrics for
binary and point forecasts. Other scoring rules have a many-to-one
relationship, returning one value per multiple values in
\texttt{predicted}. This is the case for all scoring rules for sample-
and quantile-based forecasts. For sample- and quantile-based forecasts,
\texttt{predicted} is therefore a matrix, with values in each row
jointly forming a single forecast.

Input formats and return values are shown in more detail in Figure
\ref{fig:input-scoring-rules}. The package vignettes provide extensive
documentation for the metrics exported by \pkg{scoringutils} and offer
guidance on which scoring rule to use and how to interpret the scores.

\subsection{Adding relative skill scores based on pairwise
comparisons}\label{pairwisetheory}

Raw scores for different forecasting models are usually not directly
comparable when there are missing forecasts in the data set, as
missingness is often correlated with predictive performance. One way to
mitigate this are relative skill scores based on pairwise comparisons
\citep{cramerEvaluationIndividualEnsemble2021}.

Models enter a `pairwise tournament', where all possible pairs of models
are compared based on the overlapping set of available forecasts common
to both models (omitting comparisons where there is no overlapping set
of forecasts). For every pair, the ratio of the mean scores of both
models is computed. The relative skill score of a model is then the
geometric mean of all mean score ratios which involve that model (see
Figure \ref{fig:pairwise-comparison}. This gives us an indicator of
performance relative to all other models, with the orientation depending
on the score used: if lower values are better for a particular scoring
rule, then the same is true for the relative skill score computed based
on that score.

Two models can of course only be fairly compared if they have
overlapping forecasts. Furthermore, pairwise comparisons between models
for a given score are only possible if all values have the same sign,
i.e., all score values need to be either positive or negative.

\begin{CodeChunk}
\begin{figure}[!h]

{\centering \includegraphics[width=1\linewidth]{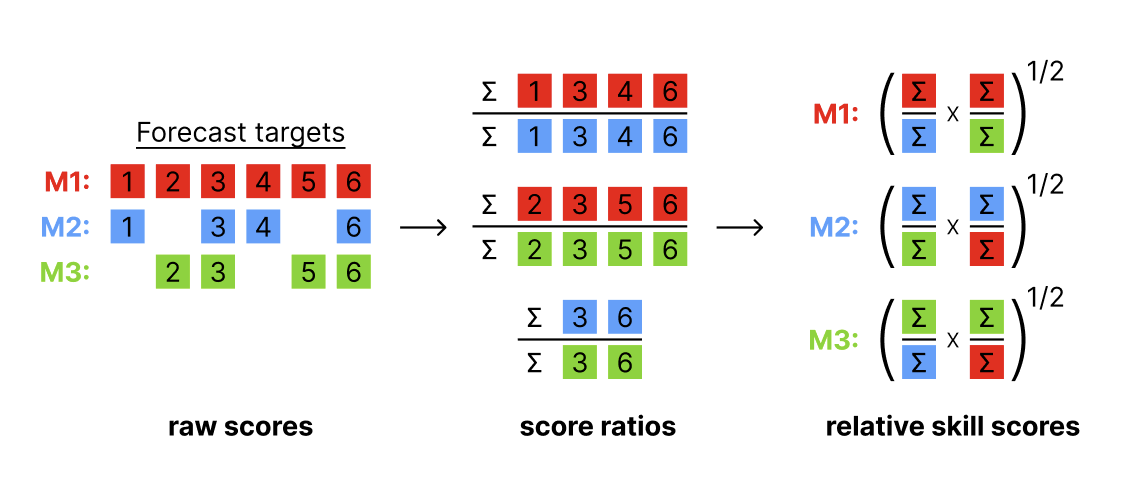} 

}

\caption[Illustration of the computation of relative skill scores through pairwise comparisons of three different forecast models, M1-M3]{Illustration of the computation of relative skill scores through pairwise comparisons of three different forecast models, M1-M3. Score ratios are computed based on the overlapping set of forecasts common to all pairs of two models. The relative skill score of a model is then the geometric mean of all mean score ratios which involve that model. The orientation of the relative skill score depends on the score used: if lower values are better for a particular scoring rule, then the same is true for the relative skill score computed based on that score.}\label{fig:pairwise-comparison}
\end{figure}
\end{CodeChunk}

To compute relative skill scores, users can call
\code{add\_pairwise\_comparison()} on the output of \code{score()}. This
function computes relative skill values with respect to a score
specified in the argument \texttt{metric} and adds them as an additional
column to the input data. Optionally, users can specify a baseline model
to also compute relative skill scores scaled with respect to that
baseline. Scaled relative skill scores are obtained by simply dividing
the relative skill score for every individual model by the relative
skill score of the baseline model. Pairwise comparisons are computed
according to the grouping specified in the argument \code{by}:
internally, the \code{data.table} with all scores gets split into
different \code{data.table}s according to the values specified in
\code{by} (excluding the column `model'). Relative scores are then
computed for every individual group separately. In the example below we
specify \code{by = c("model", "target_type")}, which means that there is
one relative skill score per model, calculated completely separately for
the different forecasting targets.

\begin{CodeChunk}
\begin{CodeInput}
R> forecast_quantile |>
+   score() |>
+   add_relative_skill(by = c("model", "target_type"), 
+                      baseline = "EuroCOVIDhub-baseline")
\end{CodeInput}
\end{CodeChunk}

Pairwise comparisons should usually be made based on raw, unsummarised
scores (meaning that \texttt{add\_relative\_skill()} should be called
before \texttt{summarise\_scores()} (see Section
\ref{sec:summarising})). Summarising scores, for example by computing an
average across several dimensions, can change the set of overlapping
forecasts between two models and distort relative skill scores.

\subsection{Additional functionality related to scores
objects}\label{additional-functionality-related-to-scores-objects}

\subsubsection{Displaying mean score ratios from pairwise
comparisons}\label{displaying-mean-score-ratios-from-pairwise-comparisons}

\pkg{scoringutils} offers a second alternative workflow to conduct
pairwise comparisons between models through the function
\texttt{get\_pairwise\_comparisons()}. The purpose of this workflow is
to obtain and visualise information on the direct comparisons between
every possible pair of models, rather than just computing relative skill
scores for every model. The function
\texttt{get\_pairwise\_comparisons()} accepts the same inputs as
\code{add\_relative\_skill()}, and returns a \code{data.table} with the
results of the pairwise tournament. These include the mean score ratios
for every pair of models, a \(p\)\textasciitilde value for whether
scores for one model are significantly different from scores for another
model, and the relative and scaled relative skill score for every model
(depending on whether a baseline was provided or not).

\texttt{get\_pairwise\_comparisons()} computes
\(p\)\textasciitilde values using either the Wilcoxon rank sum test (the
default, the test is also known as Mann-Whitney-U test)
\citep{mannTestWhetherOne1947} or a permutation test.
\(p\)\textasciitilde values are then adjusted using \texttt{p.adjust}.
In practice, the computation of \(p\)\textasciitilde values is
complicated by the fact that both tests assume independent observations.
In reality, however, forecasts by a model may be correlated across time
or space (e.g., if a forecaster has a bad day, they might perform badly
across different targets for a given forecast date).
\(p\)\textasciitilde values may therefore be too liberal in suggesting
significant differences where there aren't any. We previously suggested
computing relative skill scores based on pairwise comparisons before
summarising scores. One exception is the case where one is interested in
\(p\)\textasciitilde values specifically: One possible way to mitigate
issues from correlated forecasts, is to aggregate observations over a
category where one suspects correlation (provided there are no missing
values within the categories summarised over) to reduce correlation
before making pairwise comparisons. A test that is performed on
aggregate scores will likely be more conservative.

The mean score ratios resulting from \code{pairwise\_comparison()} can
then be visualised using the function
\code{plot\_pairwise\_comparison()}. An example is shown in Figure
\ref{fig:pairwise-plot}.

\begin{CodeChunk}
\begin{CodeInput}
R> forecast_quantile |>
+   score() |>
+   get_pairwise_comparisons(compare = "model", by = "target_type") |>
+   plot_pairwise_comparisons() + 
+   facet_wrap(~ target_type)
\end{CodeInput}
\begin{figure}

{\centering \includegraphics[width=1\linewidth]{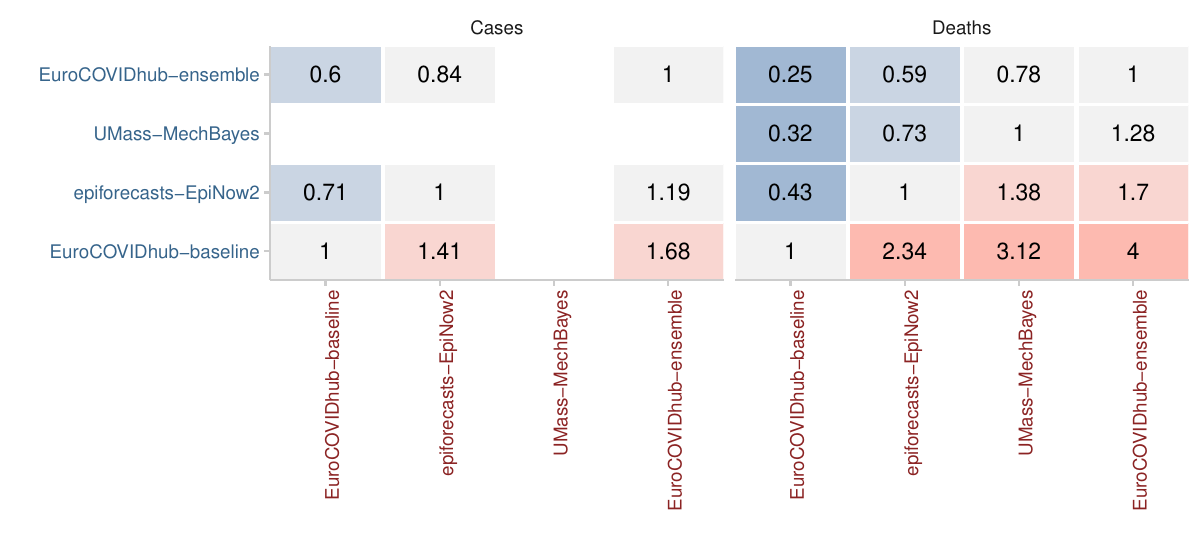} 

}

\caption[Ratios of mean weighted interval scores based on overlapping forecast sets]{Ratios of mean weighted interval scores based on overlapping forecast sets. When interpreting the plot one should look at the model on the y-axis, and the model on the x-axis is the one it is compared against. If a tile is blue, then the model on the y-axis performed better (assuming that scores are negatively oriented, i.e., that lower scores are better). If it is red, the model on the x-axis performed better in direct comparison. In the example above, the EuroCOVIDhub-ensemble performs best (it only has values smaller than one), while the EuroCOVIDhub-baseline performs worst (and only has values larger than one). For cases, the UMass-MechBayes model is excluded as there are no case forecasts available and therefore the set of overlapping forecasts is empty.}\label{fig:pairwise-plot}
\end{figure}
\end{CodeChunk}

\subsubsection{Correlations between
scores}\label{correlations-between-scores}

Users can examine correlations between scores using the function
\code{correlations()} and plot the result using
\code{plot\_correlations()}. The plot resulting from the following code
is shown in Figure \ref{fig:correlation-plot}.

\begin{CodeChunk}
\begin{CodeInput}
R> correlations <- forecast_quantile |>
+   score() |>
+   summarise_scores() |>
+   get_correlations()
R> 
R> correlations |>
+   plot_correlations(digits = 2)
\end{CodeInput}
\begin{figure}[!h]

{\centering \includegraphics[width=1\linewidth]{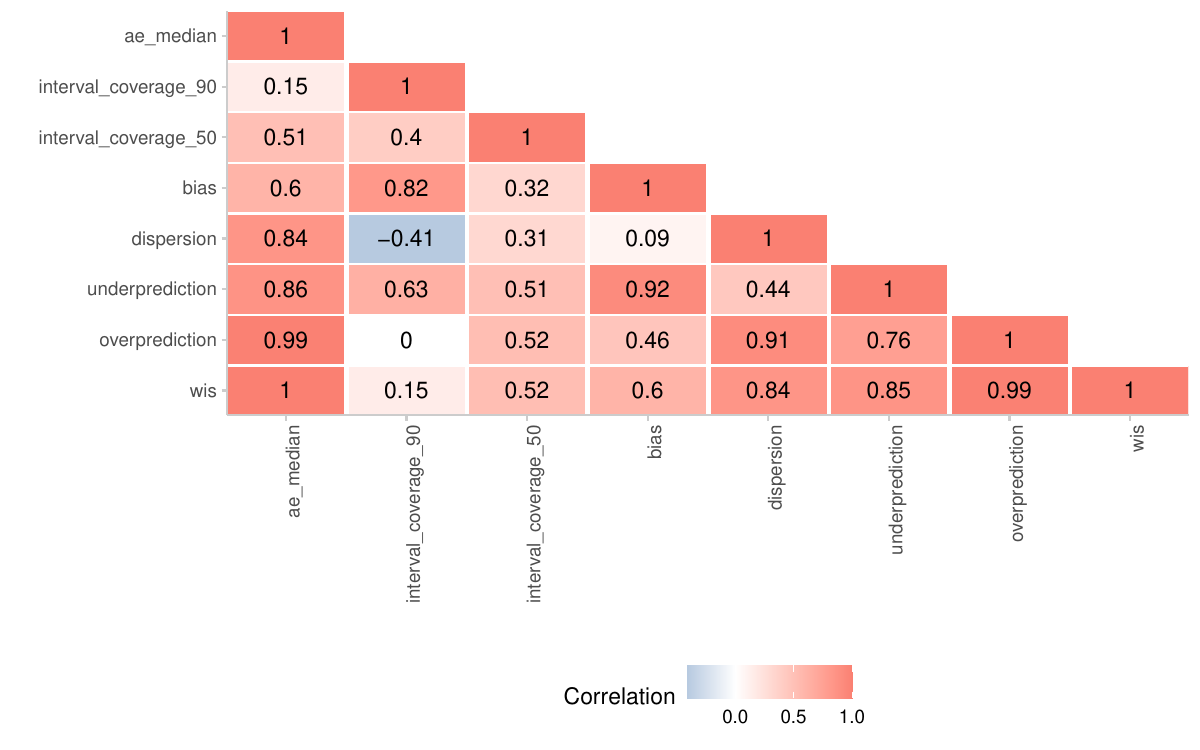} 

}

\caption[Plot of correlations between different scores]{Plot of correlations between different scores. Numbers, as well as the shade of the cells, indicate the correlation between two scores.}\label{fig:correlation-plot}
\end{figure}
\end{CodeChunk}

\section{Summarising results} \label{sec:summarising}

\subsection{Summarising scores}\label{summarising-scores}

Usually, one will not be interested in scores for each individual
forecast, but rather in summarised scores. This can be achieved using
the function \code{summarise\_scores()}. The function takes a
\texttt{scores} object (a data.table with an additional attribute
\texttt{metrics}) as input and applies a summary function to it (by
default the mean), returning a \texttt{data.table} with summarised
scores. Users can set the summary level using the argument \texttt{by}
and will obtain a summarised score for each combination of the value in
the specified columns (e.g., \texttt{by\ =\ c("model",\ "target\_type")}
will return one summarised score per model and target type). To display
scores it is often useful to round the output, for example to two
significant digits, which can be achieved with another call to
\code{summarise\_scores()}.

\begin{CodeChunk}
\begin{CodeInput}
R> forecast_quantile |>
+   score(metrics = list("wis" = wis)) |>
+   summarise_scores(by = c("model", "target_type")) |>
+   summarise_scores(fun = signif, digits = 2)
\end{CodeInput}
\begin{CodeOutput}
                   model   wis
                  <char> <num>
1: EuroCOVIDhub-ensemble 17000
2: EuroCOVIDhub-ensemble    41
3: EuroCOVIDhub-baseline 29000
4: EuroCOVIDhub-baseline   160
5:  epiforecasts-EpiNow2 21000
6:  epiforecasts-EpiNow2    69
7:       UMass-MechBayes    52
\end{CodeOutput}
\end{CodeChunk}

While \code{summarise\_scores()} accepts arbitrary summary functions,
care has to be taken when using something else than \code{mean()}, as
this may create an incentive for dishonest reporting. Many scoring rules
for probabilistic forecasts are `strictly proper scoring rules'
\citep{gneitingStrictlyProperScoring2007}, meaning that they are
constructed such that they cannot be cheated and always incentivise the
forecaster to report her honest belief about the future. Let's assume
that a forecaster's true belief about the future corresponds to a
predictive distribution \(F\). Then, if \(F\) was the true
data-generating process, a scoring rule would be proper if it ensures
that no other forecast distribution \(G\) would yield a better expected
score. If the scoring rule ensures that under \(F\) no other possible
predictive distribution can achieve the same expected score as \(F\),
then it is called strictly proper. From the forecaster's perspective,
any deviation from her true belief \(F\) leads to a worsening of
expected scores. When using summary functions other than the mean,
however, scores may lose their propriety (the property of incentivising
honest reporting) and become cheatable. For example, the median of
several individual scores (individually based on a strictly proper
scoring rule) is usually not proper. A forecaster judged by the median
of several scores may be incentivised to misrepresent their true belief
in a way that is not true for the mean score.

The user must exercise additional caution and should usually avoid
aggregating scores across categories which differ much in the magnitude
of the quantity to forecast, as (depending on the scoring rule used)
forecast errors usually increase with the order of magnitude of the
forecast target. In the given example, looking at one score per model
(i.e., specifying \code{by = c("model")}) is problematic, as overall
aggregate scores would be dominated by case forecasts, while performance
on deaths would have little influence. Similarly, aggregating over
different forecast horizons is often ill-advised as the mean will be
dominated by further ahead forecast horizons. In the previous function
calls, we therefore decided to only analyse forecasts with a forecast
horizon of two weeks.

\subsection{Additional functionality for summarised
scores}\label{additional-functionality-for-summarised-scores}

\subsubsection{Heatmaps}\label{heatmaps}

To detect systematic patterns it may be useful to visualise a single
score across several dimensions. The function \code{plot\_heatmap()} can
be used to create a heatmap that achieves this. The following produces a
heatmap of bias values across different locations and forecast targets
(output shown in Figure \ref{fig:score-heatmap}).

\begin{CodeChunk}
\begin{CodeInput}
R> example_sample_continuous[horizon == 2] |>
+   score() |>
+   summarise_scores(by = c("model", "location", "target_type")) |>
+   summarise_scores(
+     by = c("model", "location", "target_type"), 
+     fun = signif, digits = 2) |>
+   plot_heatmap(x = "location", metric = "bias") + 
+     facet_wrap(~ target_type) 
\end{CodeInput}
\begin{figure}[!h]

{\centering \includegraphics[width=1\linewidth]{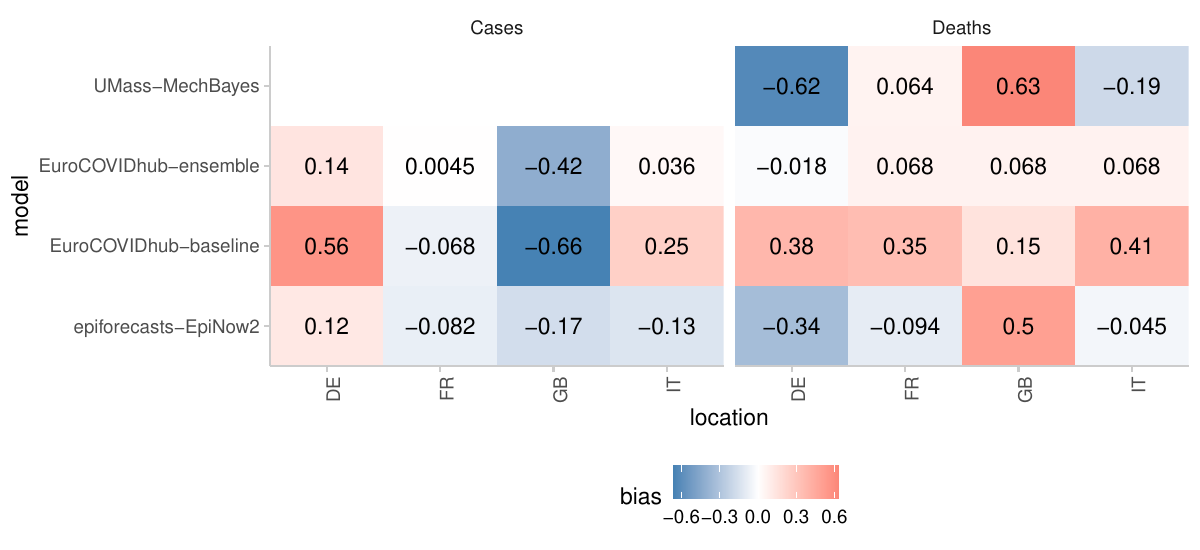} 

}

\caption[Heatmap of bias values for different models across different locations and forecast targets]{Heatmap of bias values for different models across different locations and forecast targets. Bias values are bound between -1 (underprediction) and 1 (overprediction) and should be 0 ideally. Red tiles indicate an upwards bias (overprediction), while blue tiles indicate a downwards bias (underprediction)}\label{fig:score-heatmap}
\end{figure}
\end{CodeChunk}

\subsubsection{Weighted interval score
decomposition}\label{weighted-interval-score-decomposition}

For quantile-based forecasts, the weighted interval score
\citep[WIS, ][see Section \ref{sec:wis} in the Appendix]{bracherEvaluatingEpidemicForecasts2021}
is a commonly used strictly proper scoring rule for forecasts in a
quantile-based format. The score is the sum of three components:
overprediction, underprediction and dispersion (width of the forecast).
These can be visualised using the function \code{plot\_wis()}, as shown
in Figure \ref{fig:wis-components}.

\begin{CodeChunk}
\begin{CodeInput}
R> forecast_quantile |>
+   score() |>
+   summarise_scores(by = c("model", "target_type")) |>
+   plot_wis(relative_contributions = FALSE) + 
+   facet_wrap(~ target_type, 
+              scales = "free_x") 
\end{CodeInput}
\end{CodeChunk}

\begin{CodeChunk}
\begin{figure}[!h]

{\centering \includegraphics[width=1\linewidth]{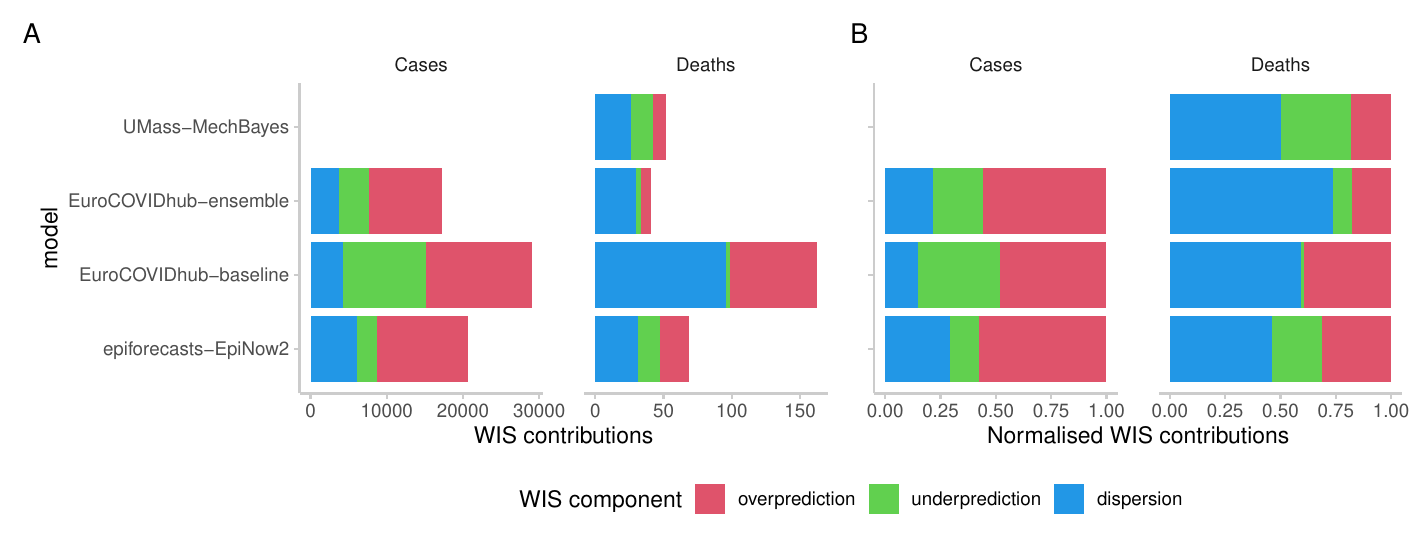} 

}

\caption[Decomposition of the weighted interval score (WIS) into dispersion, overprediction and underprediction]{Decomposition of the weighted interval score (WIS) into dispersion, overprediction and underprediction. A: absolute contributions, B: contributions normalised to 1.}\label{fig:wis-components}
\end{figure}
\end{CodeChunk}

\section{Discussion} \label{sec:discussion}

\subsubsection{Summary}\label{summary}

This paper presented \pkg{scoringutils} an R package for forecast
evaluation. It explained the core workflow, consisting of 1) validating
and processing inputs, 2) scoring forecasts and 3) summarising scores,
as well as additional functionality such as visualisation and diagnostic
tools.

The package specialises in the evaluation of probabilistic forecasts
(the forecast is a full predictive distribution). It provides a
comprehensive framework based on \pkg{data.table} and allows users to
validate, diagnose, visualise, transform and score forecasts using a
wide range of default and custom scoring rules. The package is designed
to be flexible and extensible, and to make it easy to use functionality
from different packages in a single workflow. \pkg{scoringutils}
addresses a gap in the existing ecosystem of forecast evaluation by
creating a \texttt{data.table}-based forecast evaluation framework for
probabilistic forecasts (similarly to what \pkg{yardstick} provides for
point forecasts and classification tasks). Notably, \pkg{scoringutils}
is the first package to provide extensive support for forecasts in a
quantile-based forecasts, which is commonly used for example in
Epidemiology. In addition to providing a coherent forecast evaluation
workflow it offers a wide range of additional functions that
practitioners may find useful when assessing or comparing the quality of
their forecasts.

One important limitation of the package is that it currently does not
support statistical testing of forecast performance as part of its core
workflow. Determining whether a forecaster is significantly better than
another is an important aspect of forecast evaluation that is currently
mostly missing from the package. Another limitation is the fact that the
package currently only supports a small set of possible types of
forecasts. For example, forecasts in a bin-format or forecasts
represented in a closed-form distribution (as can be scored for example
using \pkg{scoringRules} are not supported. While it is in principle
possible to extend the current classes and generic functions, this may
not be very feasible in practice for most users. Some functionality in
\pkg{scoringutils} is necessarily redundant with other packages that
provide functionality to aid with the evaluation of forecasts. The
overall idea of providing a \texttt{data.frame}-based evaluation
framework, for example, is similar to what \pkg{yardstick} offers
(albeit with a focus on point forecasts and classification tasks, rather
than probabilistic forecasts). Having a single package that encompasses
all possible use cases might be preferable. At the moment,
\pkg{scoringutils} falls somewhat short of its aspiration to become a
bridge between different packages in the forecast evaluation ecosystem.
It does not yet offer a wide range of helper functions that allow users
to easily convert between different formats and use functionality from
other packages and many visualisations that are available in other
packages, particularly with respect to model calibration, are missing.

A variety of extensions are planned for \pkg{scoringutils}. The first is
the expansion of the forecast types that are supported. We plan to add
support for evaluating categorical forecasts, as well as multivariate
forecasts that specify a joint distribution across targets. Adding the
possibility to score closed-form distributions might be another useful
extension. A second area of expansion is the integration with other
forecast evaluation and modelling packages. We aim to provide a variety
of helper functions to convert to and from different formats, such as
the one used by \pkg{yardstick} or formats used by modelling packages
such as \pkg{odin}. These functions would make it easy to integrate
\pkg{scoringutils} into existing workflows or use functionality from
other packages that is not available in \pkg{scoringutils}. A third area
of improvement is the addition of case studies and vignettes that make
working with and extending functionality from the package easier.

\pkg{scoringutils} is already used by a variety of public health
institutions such as the US Centers for Disease Control, the European
Centre for Disease Prevention and Control, as well as various academic
institutions. The package is actively maintained and developed and we
hope it will continue to be a valuable resource for researchers and
practitioners working on forecast evaluation.

\section{Acknowledgments}\label{acknowledgments}

Funding statements

NIB received funding from the Health Protection Research Unit (grant
code NIHR200908). HG's work was funded by the Wellcome Trust (grant:
210758/Z/18/Z). AC acknowledges funding by the NIHR, the Sergei Brin
foundation, USAID, and the Academy of Medical Sciences. EvL acknowledges
funding by the National Institute for Health Research (NIHR) Health
Protection Research Unit (HPRU) in Modelling and Health Economics (grant
number NIHR200908) and the European Union's Horizon 2020 research and
innovation programme - project EpiPose (101003688). SF's work was
supported by the Wellcome Trust (grant: 210758/Z/18/Z), and the NIHR
(NIHR200908). SA's work was funded by the Wellcome Trust (grant:
210758/Z/18/Z). This study is partially funded by the National Institute
for Health Research (NIHR) Health Protection Research Unit in Modelling
and Health Economics, a partnership between UK Health Security Agency
and Imperial College London in collaboration with LSHTM (grant code
NIHR200908); and acknowledges funding from the MRC Centre for Global
Infectious Disease Analysis (reference MR/R015600/1), jointly funded by
the UK Medical Research Council (MRC) and the UK Foreign, Commonwealth
\& Development Office (FCDO), under the MRC/FCDO Concordat agreement and
is also part of the EDCTP2 programme supported by the European Union.
Disclaimer: ``The views expressed are those of the author(s) and not
necessarily those of the NIHR, UKHSA or the Department of Health and
Social Care. We thank Community Jameel for Institute and research
funding. This work has also been supported by the US National Institutes
of General Medical Sciences (R35GM119582). The content is solely the
responsibility of the authors and does not necessarily represent the
official views of NIGMS, or the National Institutes of Health.

\bibliography{references.bib, scoringutils-paper.bib}

\begin{thebibliography}{42}
\newcommand{\enquote}[1]{``#1''}
\providecommand{\natexlab}[1]{#1}
\providecommand{\url}[1]{\texttt{#1}}
\providecommand{\urlprefix}{URL }
\expandafter\ifx\csname urlstyle\endcsname\relax
  \providecommand{\doi}[1]{doi:\discretionary{}{}{}#1}\else
  \providecommand{\doi}{doi:\discretionary{}{}{}\begingroup \urlstyle{rm}\Url}\fi
\providecommand{\eprint}[2][]{\url{#2}}

\bibitem[{Angus(1994)}]{angusProbabilityIntegralTransform1994}
Angus JE (1994).
\newblock \enquote{The {{Probability Integral Transform}} and {{Related Results}}.}
\newblock \emph{SIAM Review}, \textbf{36}(4), 652--654.
\newblock ISSN 0036-1445.
\newblock \doi{10.1137/1036146}.

\bibitem[{Bosse \emph{et~al.}(2023)Bosse, Abbott, Cori, van Leeuwen, Bracher, and Funk}]{bosseScoringEpidemiologicalForecasts2023}
Bosse NI, Abbott S, Cori A, van Leeuwen E, Bracher J, Funk S (2023).
\newblock \enquote{Scoring Epidemiological Forecasts on Transformed Scales.}
\newblock \emph{PLOS Computational Biology}, \textbf{19}(8), e1011393.
\newblock ISSN 1553-7358.
\newblock \doi{10.1371/journal.pcbi.1011393}.

\bibitem[{Bracher \emph{et~al.}(2021)Bracher, Ray, Gneiting, and Reich}]{bracherEvaluatingEpidemicForecasts2021}
Bracher J, Ray EL, Gneiting T, Reich NG (2021).
\newblock \enquote{Evaluating Epidemic Forecasts in an Interval Format.}
\newblock \emph{PLoS computational biology}, \textbf{17}(2), e1008618.
\newblock ISSN 1553-7358.
\newblock \doi{10.1371/journal.pcbi.1008618}.

\bibitem[{Bracher \emph{et~al.}(2022)Bracher, Wolffram, Deuschel, G{\"o}rgen, Ketterer, Ullrich, Abbott, Barbarossa, Bertsimas, Bhatia, Bodych, Bosse, Burgard, Castro, Fairchild, Fiedler, Fuhrmann, Funk, Gambin, Gogolewski, Heyder, Hotz, Kheifetz, Kirsten, Krueger, Krymova, Leith{\"a}user, Li, Meinke, Miasojedow, Michaud, Mohring, Nouvellet, Nowosielski, Ozanski, Radwan, Rakowski, Scholz, Soni, Srivastava, Gneiting, and Schienle}]{bracherNationalSubnationalShortterm2022}
Bracher J, Wolffram D, Deuschel J, G{\"o}rgen K, Ketterer JL, Ullrich A, Abbott S, Barbarossa MV, Bertsimas D, Bhatia S, Bodych M, Bosse NI, Burgard JP, Castro L, Fairchild G, Fiedler J, Fuhrmann J, Funk S, Gambin A, Gogolewski K, Heyder S, Hotz T, Kheifetz Y, Kirsten H, Krueger T, Krymova E, Leith{\"a}user N, Li ML, Meinke JH, Miasojedow B, Michaud IJ, Mohring J, Nouvellet P, Nowosielski JM, Ozanski T, Radwan M, Rakowski F, Scholz M, Soni S, Srivastava A, Gneiting T, Schienle M (2022).
\newblock \enquote{National and Subnational Short-Term Forecasting of {{COVID-19}} in {{Germany}} and {{Poland}} during Early 2021.}
\newblock \emph{Communications Medicine}, \textbf{2}(1), 1--17.
\newblock ISSN 2730-664X.
\newblock \doi{10.1038/s43856-022-00191-8}.

\bibitem[{Cramer \emph{et~al.}(2021)Cramer, Ray, Lopez, Bracher, Brennen, Rivadeneira, Gerding, Gneiting, House, Huang, Jayawardena, Kanji, Khandelwal, Le, M{\"u}hlemann, Niemi, Shah, Stark, Wang, Wattanachit, Zorn, Gu, Jain, Bannur, Deva, Kulkarni, Merugu, Raval, Shingi, Tiwari, White, Woody, Dahan, Fox, Gaither, Lachmann, Meyers, Scott, Tec, Srivastava, George, Cegan, Dettwiller, England, Farthing, Hunter, Lafferty, Linkov, Mayo, Parno, Rowland, Trump, Corsetti, Baer, Eisenberg, Falb, Huang, Martin, McCauley, Myers, Schwarz, Sheldon, Gibson, Yu, Gao, Ma, Wu, Yan, Jin, Wang, Chen, Guo, Zhao, Gu, Chen, Wang, Xu, Zhang, Zou, Biegel, Lega, Snyder, Wilson, McConnell, Walraven, Shi, Ban, Hong, Kong, Turtle, {Ben-Nun}, Riley, Riley, Koyluoglu, DesRoches, Hamory, Kyriakides, Leis, Milliken, Moloney, Morgan, Ozcan, Schrader, Shakhnovich, Siegel, Spatz, Stiefeling, Wilkinson, Wong, Gao, Bian, Cao, Ferres, Li, Liu, Xie, Zhang, Zheng, Vespignani, Chinazzi, Davis, Mu, y~Piontti, Xiong, Zheng, Baek, Farias, Georgescu, Levi, Sinha, Wilde, Penna, Celi, Sundar, Cavany, Espa{\~n}a, Moore, Oidtman, Perkins, Osthus, Castro, Fairchild, Michaud, Karlen, Lee, Dent, Grantz, Kaminsky, Kaminsky, Keegan, Lauer, Lemaitre, Lessler, Meredith, {Perez-Saez}, Shah, Smith, Truelove, Wills, Kinsey, Obrecht, Tallaksen, Burant, Wang, Gao, Gu, Kim, Li, Wang, Wang, Yu, Reiner, Barber, Gaikedu, Hay, Lim, Murray, Pigott, Prakash, Adhikari, Cui, Rodr{\'i}guez, Tabassum, Xie, Keskinocak, Asplund, Baxter, Oruc, Serban, Arik, Dusenberry, Epshteyn, Kanal, Le, Li, Pfister, Sava, Sinha, Tsai, Yoder, Yoon, Zhang, Abbott, Bosse, Funk, Hellewel, Meakin, Munday, Sherratt, Zhou, Kalantari, Yamana, Pei, Shaman, Ayer, Adee, Chhatwal, Dalgic, Ladd, Linas, Mueller, Xiao, Li, Bertsimas, Lami, Soni, Bouardi, Wang, Wang, Xie, Zeng, Green, Bien, Hu, Jahja, Narasimhan, Rajanala, Rumack, Simon, Tibshirani, Tibshirani, Ventura, Wasserman, O'Dea, Drake, Pagano, Walker, Slayton, Johansson, Biggerstaff, and Reich}]{cramerEvaluationIndividualEnsemble2021}
Cramer E, Ray EL, Lopez VK, Bracher J, Brennen A, Rivadeneira AJC, Gerding A, Gneiting T, House KH, Huang Y, Jayawardena D, Kanji AH, Khandelwal A, Le K, M{\"u}hlemann A, Niemi J, Shah A, Stark A, Wang Y, Wattanachit N, Zorn MW, Gu Y, Jain S, Bannur N, Deva A, Kulkarni M, Merugu S, Raval A, Shingi S, Tiwari A, White J, Woody S, Dahan M, Fox S, Gaither K, Lachmann M, Meyers LA, Scott JG, Tec M, Srivastava A, George GE, Cegan JC, Dettwiller ID, England WP, Farthing MW, Hunter RH, Lafferty B, Linkov I, Mayo ML, Parno MD, Rowland MA, Trump BD, Corsetti SM, Baer TM, Eisenberg MC, Falb K, Huang Y, Martin ET, McCauley E, Myers RL, Schwarz T, Sheldon D, Gibson GC, Yu R, Gao L, Ma Y, Wu D, Yan X, Jin X, Wang YX, Chen Y, Guo L, Zhao Y, Gu Q, Chen J, Wang L, Xu P, Zhang W, Zou D, Biegel H, Lega J, Snyder TL, Wilson DD, McConnell S, Walraven R, Shi Y, Ban X, Hong QJ, Kong S, Turtle JA, {Ben-Nun} M, Riley P, Riley S, Koyluoglu U, DesRoches D, Hamory B, Kyriakides C, Leis H, Milliken J, Moloney M, Morgan J, Ozcan G, Schrader C, Shakhnovich E, Siegel D, Spatz R, Stiefeling C, Wilkinson B, Wong A, Gao Z, Bian J, Cao W, Ferres JL, Li C, Liu TY, Xie X, Zhang S, Zheng S, Vespignani A, Chinazzi M, Davis JT, Mu K, y~Piontti AP, Xiong X, Zheng A, Baek J, Farias V, Georgescu A, Levi R, Sinha D, Wilde J, Penna ND, Celi LA, Sundar S, Cavany S, Espa{\~n}a G, Moore S, Oidtman R, Perkins A, Osthus D, Castro L, Fairchild G, Michaud I, Karlen D, Lee EC, Dent J, Grantz KH, Kaminsky J, Kaminsky K, Keegan LT, Lauer SA, Lemaitre JC, Lessler J, Meredith HR, {Perez-Saez} J, Shah S, Smith CP, Truelove SA, Wills J, Kinsey M, Obrecht RF, Tallaksen K, Burant JC, Wang L, Gao L, Gu Z, Kim M, Li X, Wang G, Wang Y, Yu S, Reiner RC, Barber R, Gaikedu E, Hay S, Lim S, Murray C, Pigott D, Prakash BA, Adhikari B, Cui J, Rodr{\'i}guez A, Tabassum A, Xie J, Keskinocak P, Asplund J, Baxter A, Oruc BE, Serban N, Arik SO, Dusenberry M, Epshteyn A, Kanal E, Le LT, Li CL, Pfister T, Sava D, Sinha R, Tsai T, Yoder N, Yoon J, Zhang L, Abbott S, Bosse NI, Funk S, Hellewel J, Meakin SR, Munday JD, Sherratt K, Zhou M, Kalantari R, Yamana TK, Pei S, Shaman J, Ayer T, Adee M, Chhatwal J, Dalgic OO, Ladd MA, Linas BP, Mueller P, Xiao J, Li ML, Bertsimas D, Lami OS, Soni S, Bouardi HT, Wang Y, Wang Q, Xie S, Zeng D, Green A, Bien J, Hu AJ, Jahja M, Narasimhan B, Rajanala S, Rumack A, Simon N, Tibshirani R, Tibshirani R, Ventura V, Wasserman L, O'Dea EB, Drake JM, Pagano R, Walker JW, Slayton RB, Johansson M, Biggerstaff M, Reich NG (2021).
\newblock \enquote{Evaluation of Individual and Ensemble Probabilistic Forecasts of {{COVID-19}} Mortality in the {{US}}.}
\newblock \emph{medRxiv}, p. 2021.02.03.21250974.
\newblock \doi{10.1101/2021.02.03.21250974}.

\bibitem[{Cramer \emph{et~al.}(2020)Cramer, Reich, Wang, Niemi, Hannan, House, Gu, Xie, Horstman, {aniruddhadiga}, Walraven, {starkari}, Li, Gibson, Castro, Karlen, Wattanachit, {jinghuichen}, {zyt9lsb}, {aagarwal1996}, Woody, Ray, Xu, Biegel, GuidoEspana, X, Bracher, Lee, {har96}, and {leyouz}}]{cramerCOVID19ForecastHub2020}
Cramer E, Reich NG, Wang SY, Niemi J, Hannan A, House K, Gu Y, Xie S, Horstman S, {aniruddhadiga}, Walraven R, {starkari}, Li ML, Gibson G, Castro L, Karlen D, Wattanachit N, {jinghuichen}, {zyt9lsb}, {aagarwal1996}, Woody S, Ray E, Xu FT, Biegel H, GuidoEspana, X X, Bracher J, Lee E, {har96}, {leyouz} (2020).
\newblock \enquote{{{COVID-19 Forecast Hub}}: 4 {{December}} 2020 Snapshot.}
\newblock \doi{10.5281/zenodo.3963371}.

\bibitem[{Czado \emph{et~al.}(2009)Czado, Gneiting, and Held}]{czadoPredictiveModelAssessment2009}
Czado C, Gneiting T, Held L (2009).
\newblock \enquote{Predictive {{Model Assessment}} for {{Count Data}}.}
\newblock \emph{Biometrics}, \textbf{65}(4), 1254--1261.
\newblock ISSN 1541-0420.
\newblock \doi{10.1111/j.1541-0420.2009.01191.x}.

\bibitem[{Dawid(1984)}]{dawidPresentPositionPotential1984}
Dawid AP (1984).
\newblock \enquote{Present {{Position}} and {{Potential Developments}}: {{Some Personal Views Statistical Theory}} the {{Prequential Approach}}.}
\newblock \emph{Journal of the Royal Statistical Society: Series A (General)}, \textbf{147}(2), 278--290.
\newblock ISSN 2397-2327.
\newblock \doi{10.2307/2981683}.

\bibitem[{Dowle and Srinivasan(2023)}]{data.table}
Dowle M, Srinivasan A (2023).
\newblock \emph{data.table: Extension of `data.frame`}.
\newblock R package version 1.14.8, \urlprefix\url{https://CRAN.R-project.org/package=data.table}.

\bibitem[{Elliott and Timmermann(2016)}]{elliottForecastingEconomicsFinance2016}
Elliott G, Timmermann A (2016).
\newblock \enquote{Forecasting in {{Economics}} and {{Finance}}.}
\newblock \emph{Annual Review of Economics}, \textbf{8}(1), 81--110.
\newblock \doi{10.1146/annurev-economics-080315-015346}.

\bibitem[{Epstein(1969)}]{epsteinScoringSystemProbability1969}
Epstein ES (1969).
\newblock \enquote{A {{Scoring System}} for {{Probability Forecasts}} of {{Ranked Categories}}.}
\newblock \emph{Journal of Applied Meteorology}, \textbf{8}(6), 985--987.
\newblock ISSN 0021-8952.
\newblock \doi{10.1175/1520-0450(1969)008<0985:ASSFPF>2.0.CO;2}.

\bibitem[{Funk \emph{et~al.}(2020)Funk, Abbott, Atkins, Baguelin, Baillie, Birrell, Blake, Bosse, Burton, Carruthers, Davies, Angelis, Dyson, Edmunds, Eggo, Ferguson, Gaythorpe, Gorsich, {Guyver-Fletcher}, Hellewell, Hill, Holmes, House, Jewell, Jit, Jombart, Joshi, Keeling, Kendall, Knock, Kucharski, Lythgoe, Meakin, Munday, Openshaw, Overton, Pagani, Pearson, {Perez-Guzman}, Pellis, Scarabel, Semple, Sherratt, Tang, Tildesley, {van Leeuwen}, Whittles, Group, Team, and Investigators}]{funkShorttermForecastsInform2020}
Funk S, Abbott S, Atkins BD, Baguelin M, Baillie JK, Birrell P, Blake J, Bosse NI, Burton J, Carruthers J, Davies NG, Angelis DD, Dyson L, Edmunds WJ, Eggo RM, Ferguson NM, Gaythorpe K, Gorsich E, {Guyver-Fletcher} G, Hellewell J, Hill EM, Holmes A, House TA, Jewell C, Jit M, Jombart T, Joshi I, Keeling MJ, Kendall E, Knock ES, Kucharski AJ, Lythgoe KA, Meakin SR, Munday JD, Openshaw PJM, Overton CE, Pagani F, Pearson J, {Perez-Guzman} PN, Pellis L, Scarabel F, Semple MG, Sherratt K, Tang M, Tildesley MJ, {van Leeuwen} E, Whittles LK, Group CCW, Team ICCR, Investigators I (2020).
\newblock \enquote{Short-Term Forecasts to Inform the Response to the {{Covid-19}} Epidemic in the {{UK}}.}
\newblock \emph{medRxiv}, p. 2020.11.11.20220962.
\newblock \doi{10.1101/2020.11.11.20220962}.

\bibitem[{Funk \emph{et~al.}(2019)Funk, Camacho, Kucharski, Lowe, Eggo, and Edmunds}]{funkAssessingPerformanceRealtime2019}
Funk S, Camacho A, Kucharski AJ, Lowe R, Eggo RM, Edmunds WJ (2019).
\newblock \enquote{Assessing the Performance of Real-Time Epidemic Forecasts: {{A}} Case Study of {{Ebola}} in the {{Western Area}} Region of {{Sierra Leone}}, 2014-15.}
\newblock \emph{PLOS Computational Biology}, \textbf{15}(2), e1006785.
\newblock ISSN 1553-7358.
\newblock \doi{10.1371/journal.pcbi.1006785}.

\bibitem[{Gneiting \emph{et~al.}(2007)Gneiting, Balabdaoui, and Raftery}]{gneitingProbabilisticForecastsCalibration2007}
Gneiting T, Balabdaoui F, Raftery AE (2007).
\newblock \enquote{Probabilistic Forecasts, Calibration and Sharpness.}
\newblock \emph{Journal of the Royal Statistical Society: Series B (Statistical Methodology)}, \textbf{69}(2), 243--268.
\newblock ISSN 1467-9868.
\newblock \doi{10.1111/j.1467-9868.2007.00587.x}.

\bibitem[{Gneiting and Raftery(2005)}]{gneitingWeatherForecastingEnsemble2005}
Gneiting T, Raftery AE (2005).
\newblock \enquote{Weather {{Forecasting}} with {{Ensemble Methods}}.}
\newblock \emph{Science}, \textbf{310}(5746), 248--249.
\newblock ISSN 0036-8075, 1095-9203.
\newblock \doi{10.1126/science.1115255}.

\bibitem[{Gneiting and Raftery(2007)}]{gneitingStrictlyProperScoring2007}
Gneiting T, Raftery AE (2007).
\newblock \enquote{Strictly {{Proper Scoring Rules}}, {{Prediction}}, and {{Estimation}}.}
\newblock \emph{Journal of the American Statistical Association}, \textbf{102}(477), 359--378.
\newblock ISSN 0162-1459, 1537-274X.
\newblock \doi{10.1198/016214506000001437}.

\bibitem[{Good(1952)}]{goodRationalDecisions1952}
Good IJ (1952).
\newblock \enquote{Rational {{Decisions}}.}
\newblock \emph{Journal of the Royal Statistical Society. Series B (Methodological)}, \textbf{14}(1), 107--114.
\newblock ISSN 0035-9246.
\newblock \eprint{2984087}.

\bibitem[{Hamill(2001)}]{hamillInterpretationRankHistograms2001a}
Hamill TM (2001).
\newblock \enquote{Interpretation of {{Rank Histograms}} for {{Verifying Ensemble Forecasts}}.}
\newblock \emph{Monthly Weather Review}, \textbf{129}(3), 550--560.
\newblock ISSN 1520-0493, 0027-0644.
\newblock \doi{10.1175/1520-0493(2001)129<0550:IORHFV>2.0.CO;2}.

\bibitem[{Hamner and Frasco(2018)}]{Metrics}
Hamner B, Frasco M (2018).
\newblock \emph{Metrics: Evaluation Metrics for Machine Learning}.
\newblock R package version 0.1.4, \urlprefix\url{https://CRAN.R-project.org/package=Metrics}.

\bibitem[{Jordan \emph{et~al.}(2019)Jordan, Kr\"uger, and Lerch}]{scoringRules}
Jordan A, Kr\"uger F, Lerch S (2019).
\newblock \enquote{Evaluating Probabilistic Forecasts with {scoringRules}.}
\newblock \emph{Journal of Statistical Software}, \textbf{90}(12), 1--37.
\newblock \doi{10.18637/jss.v090.i12}.

\bibitem[{Kuhn \emph{et~al.}(2023{\natexlab{a}})Kuhn, Vaughan, and Hvitfeldt}]{yardstick}
Kuhn M, Vaughan D, Hvitfeldt E (2023{\natexlab{a}}).
\newblock \emph{yardstick: Tidy Characterizations of Model Performance}.
\newblock R package version 1.2.0, \urlprefix\url{https://CRAN.R-project.org/package=yardstick}.

\bibitem[{Kuhn \emph{et~al.}(2023{\natexlab{b}})Kuhn, Vaughan, and Ruiz}]{probably}
Kuhn M, Vaughan D, Ruiz E (2023{\natexlab{b}}).
\newblock \emph{probably: Tools for Post-Processing Class Probability Estimates}.
\newblock R package version 1.0.2, \urlprefix\url{https://CRAN.R-project.org/package=probably}.

\bibitem[{Kuhn and Wickham(2020)}]{tidymodels}
Kuhn M, Wickham H (2020).
\newblock \emph{Tidymodels: a collection of packages for modeling and machine learning using tidyverse principles.}
\newblock \urlprefix\url{https://www.tidymodels.org}.

\bibitem[{Kukkonen \emph{et~al.}(2012)Kukkonen, Olsson, Schultz, Baklanov, Klein, Miranda, Monteiro, Hirtl, Tarvainen, Boy, Peuch, Poupkou, Kioutsioukis, Finardi, Sofiev, Sokhi, Lehtinen, Karatzas, San~Jos{\'e}, Astitha, Kallos, Schaap, Reimer, Jakobs, and Eben}]{kukkonenReviewOperationalRegionalscale2012}
Kukkonen J, Olsson T, Schultz DM, Baklanov A, Klein T, Miranda AI, Monteiro A, Hirtl M, Tarvainen V, Boy M, Peuch VH, Poupkou A, Kioutsioukis I, Finardi S, Sofiev M, Sokhi R, Lehtinen KEJ, Karatzas K, San~Jos{\'e} R, Astitha M, Kallos G, Schaap M, Reimer E, Jakobs H, Eben K (2012).
\newblock \enquote{A Review of Operational, Regional-Scale, Chemical Weather Forecasting Models in {{Europe}}.}
\newblock \emph{Atmospheric Chemistry and Physics}, \textbf{12}(1), 1--87.
\newblock ISSN 1680-7316.
\newblock \doi{10.5194/acp-12-1-2012}.

\bibitem[{Laboratory(2015)}]{verification}
Laboratory NRA (2015).
\newblock \emph{verification: Weather Forecast Verification Utilities}.
\newblock R package version 1.42, \urlprefix\url{https://CRAN.R-project.org/package=verification}.

\bibitem[{Liboschik \emph{et~al.}(2017)Liboschik, Fokianos, and Fried}]{tscount}
Liboschik T, Fokianos K, Fried R (2017).
\newblock \enquote{{tscount}: An {R} Package for Analysis of Count Time Series Following Generalized Linear Models.}
\newblock \emph{Journal of Statistical Software}, \textbf{82}(5), 1--51.
\newblock \doi{10.18637/jss.v082.i05}.

\bibitem[{Mann and Whitney(1947)}]{mannTestWhetherOne1947}
Mann HB, Whitney DR (1947).
\newblock \enquote{On a {{Test}} of {{Whether}} One of {{Two Random Variables}} Is {{Stochastically Larger}} than the {{Other}}.}
\newblock \emph{The Annals of Mathematical Statistics}, \textbf{18}(1), 50--60.
\newblock ISSN 0003-4851, 2168-8990.
\newblock \doi{10.1214/aoms/1177730491}.

\bibitem[{Matheson and Winkler(1976)}]{mathesonScoringRulesContinuous1976}
Matheson JE, Winkler RL (1976).
\newblock \enquote{Scoring {{Rules}} for {{Continuous Probability Distributions}}.}
\newblock \emph{Management Science}, \textbf{22}(10), 1087--1096.
\newblock ISSN 0025-1909.
\newblock \doi{10.1287/mnsc.22.10.1087}.

\bibitem[{Merkle and Steyvers(2013)}]{scoring}
Merkle EC, Steyvers M (2013).
\newblock \enquote{Choosing a Strictly Proper Scoring Rule.}
\newblock \emph{Decision Analysis}, \textbf{10}, 292--304.

\bibitem[{Meyer \emph{et~al.}(2017)Meyer, Held, and Höhle}]{surveillance}
Meyer S, Held L, Höhle M (2017).
\newblock \enquote{Spatio-Temporal Analysis of Epidemic Phenomena Using the {R} Package {surveillance}.}
\newblock \emph{Journal of Statistical Software}, \textbf{77}(11), 1--55.
\newblock \doi{10.18637/jss.v077.i11}.

\bibitem[{Murphy(1971)}]{murphyNoteRankedProbability1971a}
Murphy AH (1971).
\newblock \enquote{A {{Note}} on the {{Ranked Probability Score}}.}
\newblock \emph{Journal of Applied Meteorology and Climatology}, \textbf{10}(1), 155--156.
\newblock ISSN 1520-0450.
\newblock \doi{10.1175/1520-0450(1971)010<0155:ANOTRP>2.0.CO;2}.

\bibitem[{O'Hara-Wild \emph{et~al.}(2023)O'Hara-Wild, Hyndman, and Wang}]{fabletools}
O'Hara-Wild M, Hyndman R, Wang E (2023).
\newblock \emph{fabletools: Core Tools for Packages in the 'fable' Framework}.
\newblock R package version 0.3.4, \urlprefix\url{https://CRAN.R-project.org/package=fabletools}.

\bibitem[{{R Core Team}(2021)}]{R}
{R Core Team} (2021).
\newblock \emph{R: A Language and Environment for Statistical Computing}.
\newblock R Foundation for Statistical Computing, Vienna, Austria.
\newblock \urlprefix\url{https://www.R-project.org/}.

\bibitem[{Reich \emph{et~al.}(2019)Reich, Brooks, Fox, Kandula, McGowan, Moore, Osthus, Ray, Tushar, Yamana, Biggerstaff, Johansson, Rosenfeld, and Shaman}]{reichCollaborativeMultiyearMultimodel2019}
Reich NG, Brooks LC, Fox SJ, Kandula S, McGowan CJ, Moore E, Osthus D, Ray EL, Tushar A, Yamana TK, Biggerstaff M, Johansson MA, Rosenfeld R, Shaman J (2019).
\newblock \enquote{A Collaborative Multiyear, Multimodel Assessment of Seasonal Influenza Forecasting in the {{United States}}.}
\newblock \emph{Proceedings of the National Academy of Sciences}, \textbf{116}(8), 3146--3154.
\newblock ISSN 0027-8424, 1091-6490.
\newblock \doi{10.1073/pnas.1812594116}.

\bibitem[{Rizopoulos(2019)}]{cvGEE}
Rizopoulos D (2019).
\newblock \emph{cvGEE: Cross-Validated Predictions from GEE}.
\newblock R package version 0.3-0, \urlprefix\url{https://CRAN.R-project.org/package=cvGEE}.

\bibitem[{Rizopoulos(2023)}]{GLMMadaptive}
Rizopoulos D (2023).
\newblock \emph{GLMMadaptive: Generalized Linear Mixed Models using Adaptive Gaussian Quadrature}.
\newblock R package version 0.9-0, \urlprefix\url{https://CRAN.R-project.org/package=GLMMadaptive}.

\bibitem[{Sadatsafavi \emph{et~al.}(2023)Sadatsafavi, Safari, and Lee}]{predtools}
Sadatsafavi M, Safari A, Lee TY (2023).
\newblock \emph{predtools: Prediction Model Tools}.
\newblock R package version 0.0.3, \urlprefix\url{https://CRAN.R-project.org/package=predtools}.

\bibitem[{Sherratt \emph{et~al.}(2022)Sherratt, Gruson, Grah, Johnson, Niehus, Prasse, Sandman, Deuschel, Wolffram, Abbott, Ullrich, Gibson, Ray, Reich, Sheldon, Wang, Wattanachit, Wang, Trnka, Obozinski, Sun, Thanou, Pottier, Krymova, Barbarossa, Leith{\"a}user, Mohring, Schneider, Wlazlo, Fuhrmann, Lange, Rodiah, Baccam, Gurung, Stage, Suchoski, Budzinski, Walraven, Villanueva, Tucek, {\v S}m{\'i}d, Zaj{\'i}cek, P{\'e}rez, Reina, Bosse, Meakin, Di~Loro, Maruotti, Eclerov{\'a}, Kraus, Kraus, Pribylova, Dimitris, Li, Saksham, Dehning, Mohr, Priesemann, Redlarski, Bejar, Ardenghi, Parolini, Ziarelli, Bock, Heyder, Hotz, E., {Guzman-Merino}, Aznarte, Mori{\~n}a, Alonso, {\'A}lvarez, L{\'o}pez, Prats, Burgard, Rodloff, Zimmermann, Kuhlmann, Zibert, Pennoni, Divino, Catal{\`a}, Lovison, Giudici, Tarantino, Bartolucci, Jona, Mingione, Farcomeni, Srivastava, {Montero-Manso}, Adiga, Hurt, Lewis, Marathe, Porebski, Venkatramanan, Bartczuk, Dreger, Gambin, Gogolewski, {Gruziel-Slomka}, Krupa, Moszynski, Niedzielewski, Nowosielski, Radwan, Rakowski, Semeniuk, Szczurek, Zielinski, Kisielewski, Pabjan, Holger, Kheifetz, Scholz, Bodych, Filinski, Idzikowski, Krueger, Ozanski, Bracher, and Funk}]{sherrattPredictivePerformanceMultimodel2022}
Sherratt K, Gruson H, Grah R, Johnson H, Niehus R, Prasse B, Sandman F, Deuschel J, Wolffram D, Abbott S, Ullrich A, Gibson G, Ray {\relax EL}, Reich {\relax NG}, Sheldon D, Wang Y, Wattanachit N, Wang L, Trnka J, Obozinski G, Sun T, Thanou D, Pottier L, Krymova E, Barbarossa {\relax MV}, Leith{\"a}user N, Mohring J, Schneider J, Wlazlo J, Fuhrmann J, Lange B, Rodiah I, Baccam P, Gurung H, Stage S, Suchoski B, Budzinski J, Walraven R, Villanueva I, Tucek V, {\v S}m{\'i}d M, Zaj{\'i}cek M, P{\'e}rez {\'A}C, Reina B, Bosse {\relax NI}, Meakin S, Di~Loro A, Maruotti A, Eclerov{\'a} V, Kraus A, Kraus D, Pribylova L, Dimitris B, Li {\relax ML}, Saksham S, Dehning J, Mohr S, Priesemann V, Redlarski G, Bejar B, Ardenghi G, Parolini N, Ziarelli G, Bock W, Heyder S, Hotz T, E SD, {Guzman-Merino} M, Aznarte {\relax JL}, Mori{\~n}a D, Alonso S, {\'A}lvarez E, L{\'o}pez D, Prats C, Burgard {\relax JP}, Rodloff A, Zimmermann T, Kuhlmann A, Zibert J, Pennoni F, Divino F, Catal{\`a} M, Lovison G, Giudici P, Tarantino B, Bartolucci F, Jona LG, Mingione M, Farcomeni A, Srivastava A, {Montero-Manso} P, Adiga A, Hurt B, Lewis B, Marathe M, Porebski P, Venkatramanan S, Bartczuk R, Dreger F, Gambin A, Gogolewski K, {Gruziel-Slomka} M, Krupa B, Moszynski A, Niedzielewski K, Nowosielski J, Radwan M, Rakowski F, Semeniuk M, Szczurek E, Zielinski J, Kisielewski J, Pabjan B, Holger K, Kheifetz Y, Scholz M, Bodych M, Filinski M, Idzikowski R, Krueger T, Ozanski T, Bracher J, Funk S (2022).
\newblock \enquote{Predictive Performance of Multi-Model Ensemble Forecasts of {{COVID-19}} across {{European}} Nation.}
\newblock \emph{Europe PMC}.
\newblock \doi{10.1101/2022.06.16.22276024}.

\bibitem[{Siegert(2020)}]{SpecsVerification}
Siegert S (2020).
\newblock \emph{SpecsVerification: Forecast Verification Routines for Ensemble Forecasts of Weather and Climate}.
\newblock R package version 0.5-3, \urlprefix\url{https://CRAN.R-project.org/package=SpecsVerification}.

\bibitem[{Timmermann(2018)}]{timmermannForecastingMethodsFinance2018}
Timmermann A (2018).
\newblock \enquote{Forecasting {{Methods}} in {{Finance}}.}
\newblock \emph{Annual Review of Financial Economics}, \textbf{10}(1), 449--479.
\newblock \doi{10.1146/annurev-financial-110217-022713}.

\bibitem[{Yan(2016)}]{MLmetrics}
Yan Y (2016).
\newblock \emph{MLmetrics: Machine Learning Evaluation Metrics}.
\newblock R package version 1.1.1, \urlprefix\url{https://CRAN.R-project.org/package=MLmetrics}.

\bibitem[{Zeileis and Lang(2022)}]{topmodels}
Zeileis A, Lang MN (2022).
\newblock \emph{topmodels: Infrastructure for Inference and Forecasting in Probabilistic Models}.
\newblock R package version 0.1-0/r1498, \urlprefix\url{https://R-Forge.R-project.org/projects/topmodels/}.

\end{thebibliography}

\clearpage

\appendix
\renewcommand\thefigure{\thesection.\arabic{figure}}

\section{Constructing and validating forecast objects}

The following section gives an overview of how \pkg{scoringutils}
constructs forecast objects. The \texttt{forecast} class comes with a
constructor, \code{new\_forecast()}, a generic validation function,
\code{assert\_forecast()}, and a convenient wrapper function
\code{as\_forecast\_...()}.

\code{new\_forecast()} constructs a \texttt{forecast} object based on a
\texttt{data.frame} or similar. It makes a deep copy of the input and
converts it into a \texttt{data.table}, adds a \texttt{model} column
with value ``Unspecified model'' if there isn't one and adds a class
\texttt{forecast\_*}, where \texttt{*} depends on the forecast type to
the object.

\code{assert\_forecast()} is a generic which dispatches to a specialised
validator method depending on the class of the input. It validates the
input and returns it if it is valid. If the input is not valid, it
throws an error with a message that explains what went wrong.

\code{as\_forecast\_...()} (optionally) renames existing columns to
conform with the requirements for forecast objects, (optionally) sets
the forecast unit, constructs the class and validates the input.

\clearpage

\section{Comparing different calibration plots}

The following Figure gives a more detailed overview of how to interpret
different calibration plots (showing the actual forecasts and
observations that produced the corresponding visualisations).

\begin{CodeChunk}
\begin{CodeOutput}

          observed    id   predicted sample_id           model
             <num> <int>       <num>     <int>          <fctr>
       1: 1.648182     1 -0.91277715         1   Pred: N(0, 1)
       2: 1.648182     1 -0.47014373         2   Pred: N(0, 1)
       3: 1.648182     1 -1.01383996         3   Pred: N(0, 1)
       4: 1.648182     1 -0.82421072         4   Pred: N(0, 1)
       5: 1.648182     1 -0.27920773         5   Pred: N(0, 1)
      ---                                                     
15999996: 1.780756  2000  0.07610071      1996 Pred: N(0, 0.5)
15999997: 1.780756  2000  0.01134733      1997 Pred: N(0, 0.5)
15999998: 1.780756  2000 -0.02880482      1998 Pred: N(0, 0.5)
15999999: 1.780756  2000  0.05465219      1999 Pred: N(0, 0.5)
16000000: 1.780756  2000 -0.17624218      2000 Pred: N(0, 0.5)
\end{CodeOutput}
\begin{figure}[!h]

{\centering \includegraphics[width=1\linewidth,]{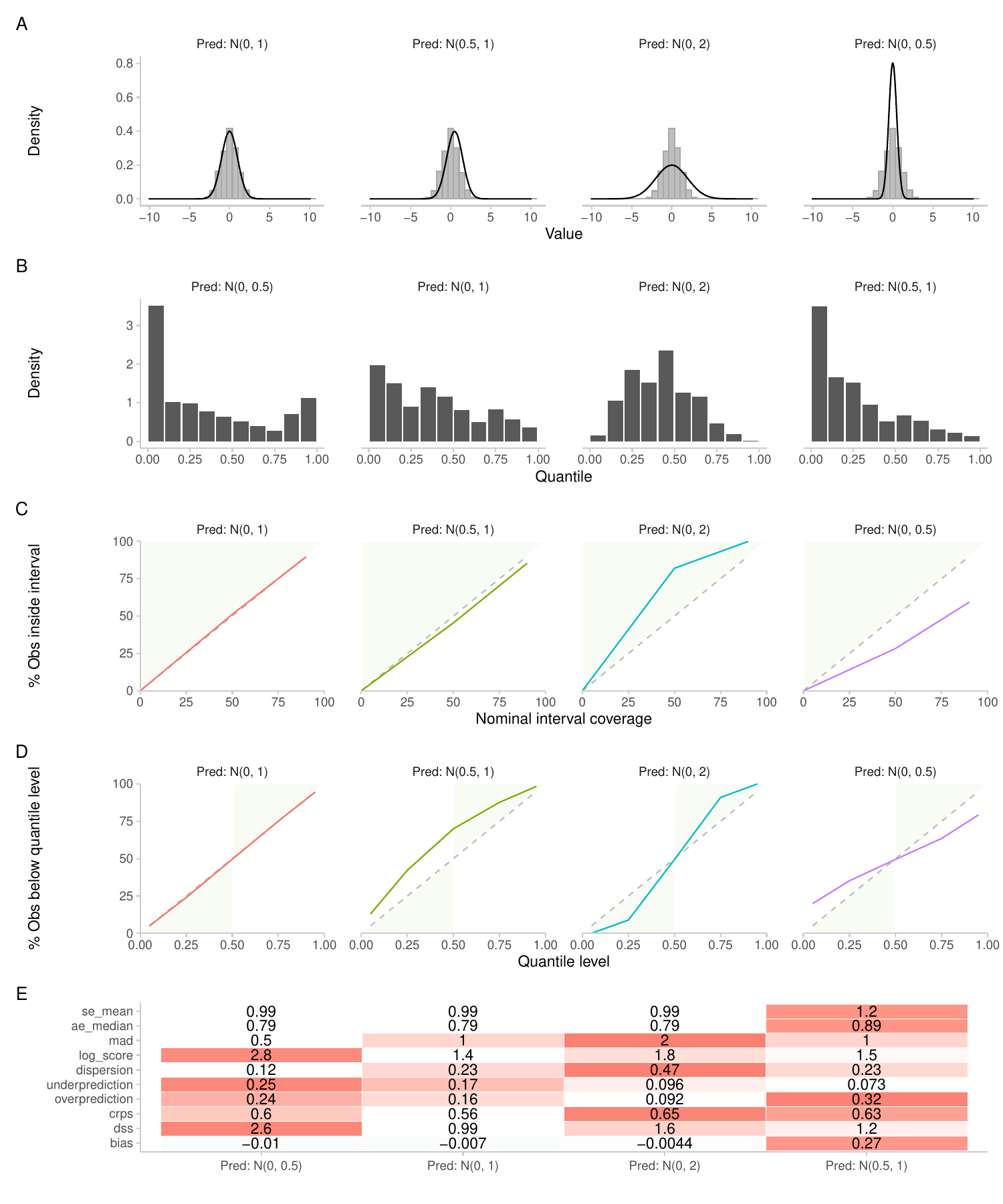} 

}

\caption[A]{A: Different forecasting distributions (black) against observations sampled from a standard normal distribution (grey histograms). B: PIT histograms based on the predictive distributions and the sampled observations shown in A. C: Empirical vs. nominal coverage of the central prediction intervals for simulated observations and predictions. Areas shaded in green indicate that the forecasts are too wide (i.e., underconfident), covering more true values than they actually should, while areas in white indicate that the model generates too narrow predictions and fails to cover the desired proportion of true values with its prediction intervals. D: Quantile coverage values, with green areas indicating too wide (i.e., conservative) forecasts. E: Scores for the standard normal predictive distribution and the observations drawn from different data-generating distributions.}\label{fig:calibration-plots}
\end{figure}
\end{CodeChunk}

\clearpage

\section{Details on the weighted interval score (WIS)} \label{sec:wis}

The WIS treats the predictive quantiles as a set of symmetric prediction
intervals and measures the distance between the observation and the
forecast interval. It can be decomposed into a dispersion (uncertainty)
component and penalties for over- and underprediction. For a single
interval, the interval score is computed as
\[IS_\alpha(F,y) = \underbrace{(u-l)}_\text{dispersion} + \underbrace{\frac{2}{\alpha} \cdot (l-y) \cdot \mathbf{1}(y \leq l)}_{\text{overprediction}} + \underbrace{\frac{2}{\alpha} \cdot (y-u) \cdot \mathbf{1}(y \geq u)}_{\text{underprediction}}, \]
where \(\mathbf{1}()\) is the indicator function, \(y\) is the observed
value, and \(l\) and \(u\) are the \(\frac{\alpha}{2}\) and
\(1 - \frac{\alpha}{2}\) quantiles of the predictive distribution \(F\),
i.e., the lower and upper bound of a single prediction interval. For a
set of \(K\) prediction intervals and the median \(m\), the score is
computed as a weighted sum,
\[WIS = \frac{1}{K + 0.5} \cdot \left(w_0 \cdot |y - m| + \sum_{k = 1}^{K} w_k \cdot IS_{\alpha}(F, y)\right),\]
where \(w_k\) is a weight for every interval. Usually,
\(w_k = \frac{\alpha_k}{2}\) and \(w_0 = 0.5\).

\clearpage

\end{document}